\documentclass[[authoryear,10pt]{elsarticle}
\pdfoutput=1
\usepackage{graphicx}
\usepackage{amssymb,amsmath}
\usepackage{bm}
\usepackage{rotating}
\usepackage{nicefrac}
\usepackage[mathlines,left]{lineno}
\usepackage{enumerate}

\begin{document}
%
\begin{frontmatter} 
\title{Effects of liquid pore water on acoustic wave propagation in snow as a Biot-type porous material}
\author[sfu]{Rolf Sidler}
\ead{rsidler@gmail.com}
\address[sfu]{Department of Earth Sciences, Simon Fraser University, 8888 University Drive, BC V5A1S6 Burnaby,
Canada}
\journal{Cold Regions Science and Technology}

\begin{abstract}

A method to estimate phase velocity and attenuation of acoustic waves in the presence of liquid water in a snowpack is presented. The method is based on Biot's theory of wave propagation in porous materials. Empirical relations and {\it a priori} information is used to characterize snow as a porous material as a function of porosity. Plane wave theory and an equivalent pore fluid are used to solve Biot's differential equations and to asses the impact of the air and water in the pore space.
The liquid water in the pore space of a snow pack reduces the velocity of the first compressional wave by roughly 300 m/s for every 0.1 increase in liquid water saturation. Also the attenuation of the compressional waves is increased with increasing liquid water content.
Two end member models for compaction are evaluated to asses the importance of an independent density measurement for an estimate of liquid pore water saturation in snow with acoustic waves.
The two end members correspond to no compaction at all and to a melting sphere packing where the grains remain in contact.
The change of velocity for the first compressional wave was found to strongly depend on the compaction model, while the velocity of the second compressional wave mainly depends on liquid water saturation. Also the attenuation for both compressional waves depends mainly on the liquid water saturation and only little on the compaction model.
Finally, a simple field experiment illustrates the potential use of the method to estimate the liquid water content of snow with acoustic sensors.
\end{abstract}

\begin{keyword}
snow \sep acoustic wave propagation \sep liquid water  \sep porous material 
\end{keyword}

\end{frontmatter}

\section{Introduction}

Knowledge of the liquid water content in snow is important to understand and forecast the storage and transmission rates of melt water or rain in a snow covered water shed \citep{gerdel:1954,colbeck:1979}. 
Also the snow albedo depends on the liquid water content and is important for remote sensing and energy balance modeling \citep{gupta:2005}.
This is due to the large dielectric contrast between liquid water and ice that increases attenuation of electromagnetic waves in the presence of liquid water. 
But the reflective behavior also depends strongly on the geometry of the pore space, which consists of larger grain sizes in wet or refrozen snow \citep{shi:1995}.
In fact, also wet snow avalanches and the corresponding stability of wet snowpacks are thought to depend on liquid water content, but the governing mechanisms are not well understood \citep{conway:1993,schweizer:2003}.

Today, mostly electromagnetic methods are used to measure the liquid water content of snow in place \citep{denoth:1984,denoth:1989,marshall:2008,techel:2011}.
Surprisingly, no attempts have been reported to measure the liquid water content in snow with acoustic waves. 

Acoustic waves have been recognized as an important tool to characterize snow and acoustic wave velocities as potential index property for snow \citep{mellor:1975}.
However, seemingly enigmatic observations as, for example, decreasing wave velocities with increasing snow density made it difficult to set acoustic observations in context to other mechanical observations for snow \citep{oura:1952,ishida:1965}.
Although \citet{johnson:1982} and \citet{sommerfeld:1982} showed that Biot's \citeyearpar{biot:1956,biot:1956a} theory for wave propagation in porous materials is well suited to predict acoustic velocities in snow and can also explain these observations of `slow' compressional waves, the application of acoustic waves for probing snow has remained sparse.

A limitation for the application of Biot's theory of wave propagation in porous materials is the need for a large number of properties that characterize the material and which are difficult to collect for snow under field conditions. Recently, \citet{sidler:2014a} 
used empirical relationships, {\it a priori} information and geometrical considerations to express snow as a Biot-type porous materials as a function of porosity. This model eliminates the need for a profound laboratory snow analysis for the application of porous wave propagation theory at the cost of, yet to be quantified, accuracy of the predicted results.
 
Porous wave propagation theory has historically been of interest for hydrocarbon exploration and more recently also for hydrogeological applications and non destructive testing \citep{mavko:1998,tang:2004,pride:2005}.
It is well known, that a mix of gas and fluid in the pore space significantly alters the wave propagation in a porous material and therefor can be used to quantify their presence \citep{mavko:1979,muller:2010}.

Wave phenomena tend to be sensitive to average material properties within a fraction of the dominant wave length. By varying the frequency of an experiment, acoustic waves can be used to investigate average liquid water content from the centimeter scale up to several meters. 
Therefore, the use of acoustic waves for the characterization of liquid water in snow has the advantage that the range of applicable wave lengths spans over the whole range of interest for spatial heterogeneity of liquid water in a snow pack.
 A characteristic that might proof to be useful to investigate the strongly heterogeneous distributions of liquid water commonly observed in natural snowpacks \citep{techel:2011}.
Possible applications may also include a combination of acoustic and electromagnetic measurements for higher precision without the need for calibration.


\section{Methods}

\subsection{Porous material properties of snow}

For the stress strain relationships in Biot's \citeyearpar{biot:1956} theory of wave propagation in porous materials four  properties of a material need to be measured in addition to the porosity, which were named $A$, $N$, $Q$, and $R$.
\citet{biot:1957} then introduced the poroelastic incompressibility $M$ and the effective stress coefficient $\alpha$ to express the porous material properties in terms of the Lam\'e constants $\lambda$ and $\mu$ and the jacketed and unjacketed compressibility coefficients $\kappa$ and $\delta$.
\citet{biot:1962} then expressed the constants in terms of the bulk modulus of the frame material $K_S$, the bulk modulus of the solid matrix $K_m$, the shear modulus $\mu$, and the fluid bulk modulus $K_f$ in the form which is generally used today.
For the equations of motion the densities of the solid and fluid material $\rho_s$ and $\rho_f$, the porosity $\phi$, and the tortuosity $\mathcal{T}$ of the porous material need to be known.
The energy dissipation due to the relative motion of the fluid to the solid based on Darcy's \citeyearpar{darcy:1856} law depends on the permeability $\kappa$ of the material and the viscosity $\nu$ of the pore fluid.
 
Therefore, seven properties define the porous frame and three define the fluid in the pore space.
However, the individual properties are not independent of each other, but are based on the structure of the frame. 
Empirical relationships, {\it a priori} information, and geometrical considerations can be used to relate the matrix properties to each other.
Snow is particularly well suited for such a relation because the frame matrix and pore fluids are relatively pure. 
Ice which behaves almost elastic for small strains is expected to be well suited for basic Biot theory where no viscoelastic effects are considered for the frame matrix \citep{gold:1958}. 
For the purpose of this study we use the relations of \citet{sidler:2014a}
to express the porous properties of snow as a function of porosity. An overview of the relationships that specify snow as a Biot-type porous material as a function of porosity are shown in Table \ref{tab:snowmodel}.

\begin{table}
\caption{Porous frame material properties for a Biot-type snow model as a function of porosity \citep{sidler:2014a}. }
\begin{center}
\begin{tabular}{lll}
\hline
Snow frame properties & \\
\hline
 	frame bulk modulus	& $K_{\rm S}$ (GPa) 	& 10  \\
 	matrix bulk modulus	& $K_{\rm m}$ (GPa)	& $K_{\rm S} (1-\phi)^\frac{30.85}{(7.76-\phi)}$ \\
 	shear modulus		& $\mu_{\rm S}$ (GPa)	& $  \frac{3}{2} \frac{K_{\rm m} (1-2 \nu)}{1+\nu}$ \\
	density			& $\rho$ (kg/m$^3$)		& $(1-\phi) \cdot 916.7 $ \\
	permeability		& $\kappa$ (m$^2$)		& $0.2 \frac{\phi^3}{(\text{SSA})^2 (1-\phi)^2}$ \\ 
	tortuosity			& ${\cal T}$			& $\frac{1}{2} \left( 1+\frac{1}{\phi} \right)$ \\
\hline
\end{tabular}
\end{center}
\label{tab:snowmodel}
\end{table}%

\subsection{Equivalent pore fluid}

In wet snow the pore space is filled with a mix of air and water. 
To model the physics of wave propagation correctly, additional field variables would have to be introduced to extend Biot's theory and account for the second phase in the pore space. Similar to the extension in the Double Porosity Model of \citet{pride:2003,pride:2003a}. 
However, as a first order approximation it is common practice to define an effective upscaled pore fluid based on the assumption that one of the phases is embedded in the other phase \citep{pride:2003,carcione:2006b}.
As in snow there is only one frame matrix material and the pore space is not separated into two distinctly different pore regimes with individual porosity, tortuosity and permeability. The use of an upscaled pore fluid seems to be a valid approach also in this case.
The effective viscosity of the pore fluid can be computed as
\begin{equation}
\label{eq:effvisc}
\eta_{\rm eff} = \eta_{\rm \textsc{a}} \left( \frac{\eta_{\rm \textsc{w}} }{ \eta_{\rm \textsc{a}}} \right )^ {S_{\rm \textsc{w}}},
\end{equation}
where $ \eta_{\rm a}$ and $\eta_{\rm w}$ are the viscosity of air and water and $S_{\rm w}$ and $S_{\rm a} = 1- S_{\rm w}$ are the fraction of water and air in the pore space, respectively \citep{teja:1981}. 
The effective bulk modulus can be obtained according to the Wood equation \citep{wood:1955,mavko:2009}
\begin{equation}
\label{eq:effmod}
K_{\rm eff}  =  \left( \frac{S_{\rm \textsc{a}} }{ K_{\rm  \textsc{a}} }+ \frac{S_{\rm \textsc{w}} }{ K_{\rm \textsc{w}}} \right) ^ {-1}, 
\end{equation}
and the effective density is
\begin{equation}
\label{eq:effdens}
\rho_{\rm eff} =  S_{\rm \textsc{a}} \rho_{\rm \textsc{a}} + S_{\rm \textsc{w}} \rho_{\rm \textsc{w}} .
\end{equation}
The values for bulk modulus, density, and viscosity for water and air used in this study are given in Table \ref{tab:porefluid}.

\begin{table}
\caption{Pore fluid properties for water and air.}
\begin{center}
\begin{tabular}{lll}
\hline
Air & \\
\hline
	density		& $\rho_{\rm a}$ (kg/m$^{3}$)	& 1.29 \\
 	viscosity		& $\eta_{\rm a}$ (Pa s)		& $1.7 \cdot 10^{-5}$  \\
	bulk modulus	& $K_{\rm a}$ (Pa)			& $1.4 \cdot 10^{5}$  \\
\hline
Water & \\
\hline
	density		& $\rho_{\rm a}$ (kg/m$^{3}$)	& 1000 \\
 	viscosity		& $\eta_{\rm a}$ (Pa s)		& $1.05 \cdot 10^{-3}$  \\
	bulk modulus	& $K_{\rm a}$ (Pa)			& $2.25 \cdot 10^{9}$  \\
\hline

\end{tabular}
\end{center}
\label{tab:porefluid}
\end{table}%

The following steps then lead to an estimate of acoustic wave velocities and attenuation as a function of liquid water saturation and density of a snowpack:
\begin{enumerate}[(i)]
\item Compute porosity of the snowpack from the snow density and the liquid pore water saturation.
\item Compute frame and pore fluid properties based on Table \ref{tab:snowmodel} and \ref{tab:porefluid} and equations (\ref{eq:effvisc}), (\ref{eq:effmod}), and (\ref{eq:effdens}).
\item Solve Biot's equations for the obtained properties with a plane wave approach (\ref{apx:planewave}). 
\end{enumerate}

As the density of liquid water is comparable to the density of ice, the mass of the pore fluid can not be neglected in wet snow.
The density $\rho$ of the entire snowpack is then given by  
\begin{equation}
\label{eq:density}
\rho  = (1-\phi) \rho_{\rm I} + \phi \left[ (1-S_{\rm w}) \rho_{\rm A} + S_{\rm w} \rho_{\rm W} \right],
\end{equation}
where $\rho_{\rm I}$, $\rho_{\rm A}$, and $\rho_{\rm W}$ are the densities of ice, air and water, respectively. 
As a first order approximation, as in dry snow, the density of the air in the pore space $\phi (1-S_{\rm w}) \rho_{\rm A}$ can be neglected as it is orders of magnitude smaller than the other densities.

\begin{figure}
\centering
\includegraphics[width=80mm]{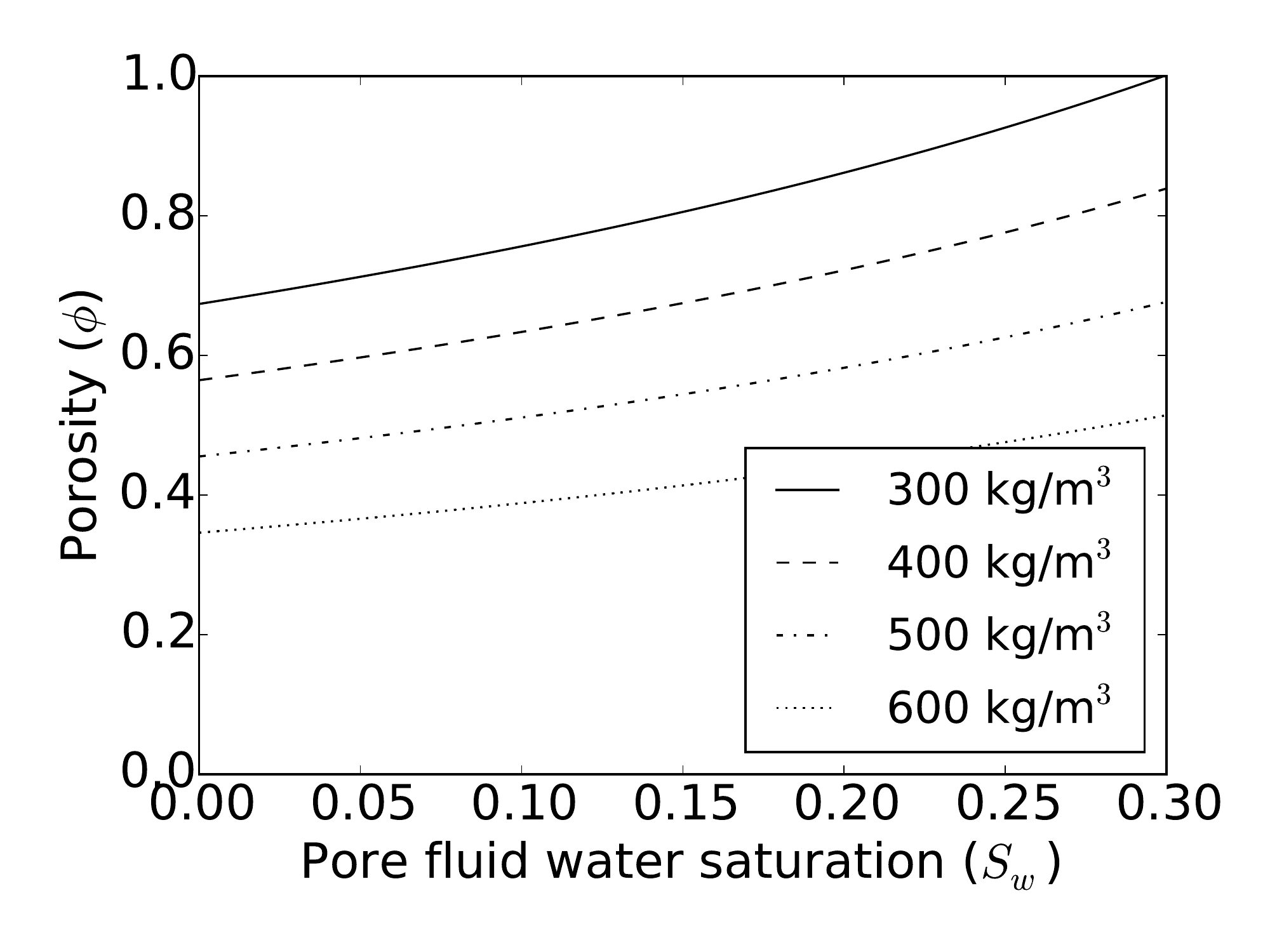}
\caption{\label{fig:por4sat} The porosity in melting snow is increasing as the liquid water is accounted to the pore space. 
Porosity change as a function of pore water saturation is shown under the assumption of mass and volume conservation for snowpacks with different dry densities.}
\end{figure}

Given the density of the snow and the liquid water saturation in the pore space the porosity can be estimated as
\begin{equation}
\label{eq:porchange}
\phi (\rho, S_{\rm \textsc{w}}) = \frac{\rho -\rho_{\rm \textsc{i}}}{\rho_{\rm eff} -\rho_{\rm \textsc{i}}},
\end{equation}
where $\rho_{\rm eff}$ can be replaced with $S_{\rm \textsc{w}} \rho_{\rm \textsc{w}}$ as a first order approximation.
The change of porosity with increasing water saturation for a constant density of the snowpack is shown in Figure \ref{fig:por4sat}.

\subsection{End-member models for snow compaction}

Two end-member models for snow compaction during snow melt are presented to predict how acoustic measurements would change for a fixed installation in a snowpack over a period of snowmelt and to asses the need of an independent measurement of snow density in an acoustic measurement to estimate the liquid water content.
Existing snowpack models reproduce the complex interactions between meteorological conditions, geothermal heat flow and the corresponding response of the snow \citep{colbeck:1982,jordan:1991,lynch:1994}.
For this study much simpler models are used with the only objective to assess a realistic upper and lower bound for the acoustic response to an increase of liquid water content in the pore space.

With the introduction of liquid water in the porous snow model shown in Table \ref{tab:snowmodel}, not only the pore fluid properties change, but also the assumption that the density of the porous frame is also the density of the entire snowpack is no longer valid. Equation (\ref{eq:porchange}) illustrates how this second degree of freedom is introduced beneath the pore water saturation.
Using the end-member models the variation is limited to a band of plausible values. Therefore reducing the complexity of the analysis and leading to more comprehensive results without loss of generality.
Moreover, if a specific property of the acoustic experiments shows equal variation with liquid water saturation, this is an indication that this property is sensitive to pore water saturation only, eliminating the need for an independent density measurement.

\begin{figure}[h!]
\centering
\includegraphics[angle=0,width=60mm]{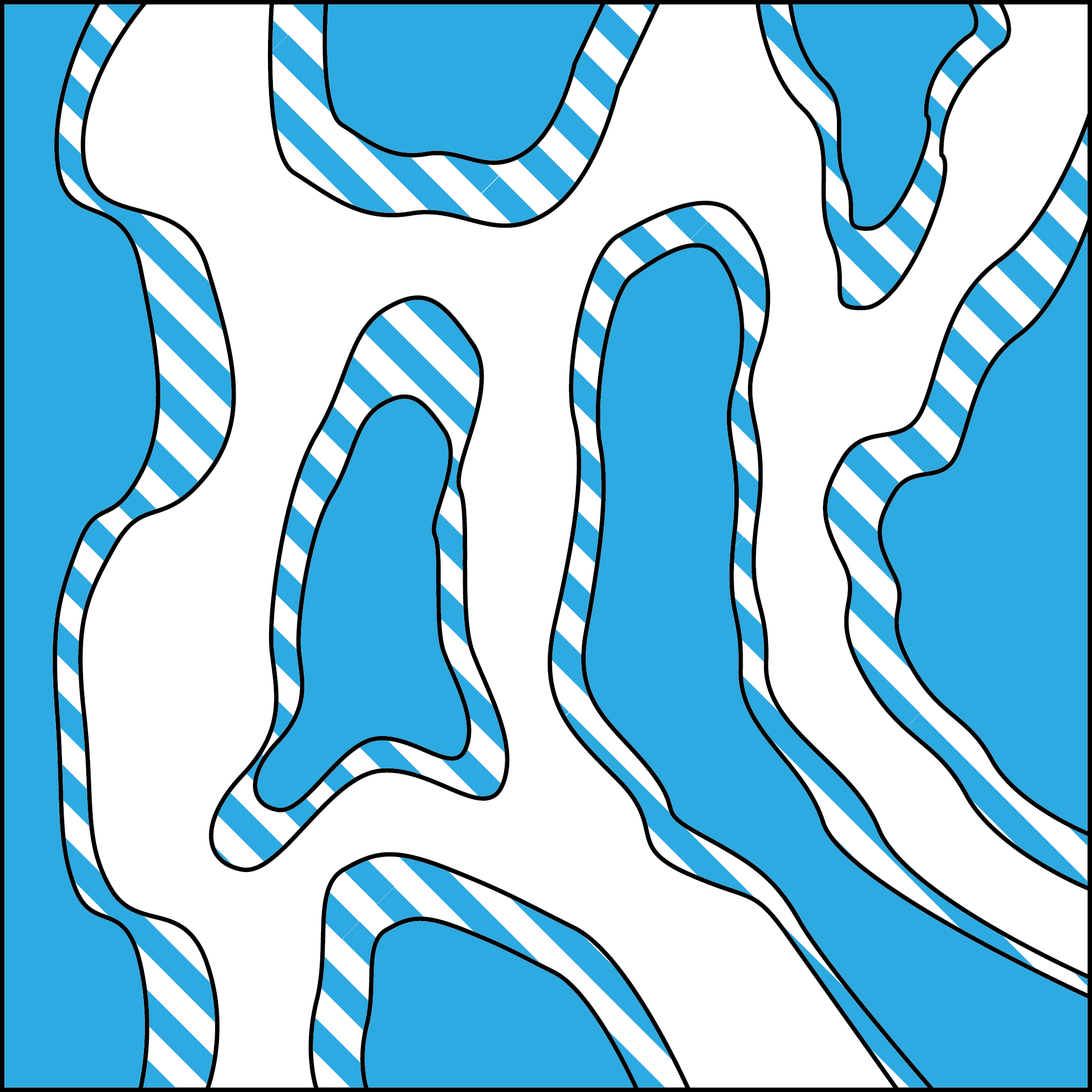} 
\caption[]{\label{fig:const-dens} End-member model for constant density. Schematic illustration of the increase of porosity in melting snow due to accounting melted parts of the ice frame to pore space instead of frame. The dashed area depicts ice of the pore frame that has melted and is consequently accounted to the pore space due to its ability to move against the porous frame.}
\end{figure}

The first end-member model assumes no snow compaction at all. It is also assumes that no liquid water is drained from the snowpack and therefore the density of the snowpack remains constant with increasing liquid water saturation. This end-member also corresponds to the situation were a single acoustic measurement on a wet snowpack is performed and evaluated against an independent density measurement.
In this model the porosity increases with increasing water saturation $S_{\rm \textsc{w}}$ following equation (\ref{eq:porchange}).
The porosity increases because the part of the ice frame that has melted is accounted to the pore space. This is necessary as the liquid water is capable to move against the porous frame.
This increase of porosity due to snowmelt for constant density is schematically illustrated in Figure \ref{fig:const-dens}.
The striped area depicts the ice that has melted and is now part of the pore space. The pore space occupies now a larger fraction of the volume.
 
There is a strong impact on wave propagation from this end-member. As the porosity changes, most of the properties of the porous frame change according to the relationships in Table \ref{tab:snowmodel}.
In addition also the equivalent pore fluid changes with the increase of the liquid water saturation. 
The use as an end-member is justified by the fact, that a stronger decrease of porosity would involve an increase of the total volume of the snow.

\begin{figure}
\centering
\begin{minipage}[t]{3mm}
\large a) 
\end{minipage}
\begin{minipage}[t]{60mm}
\vspace{-10pt}
\includegraphics[width=\textwidth]{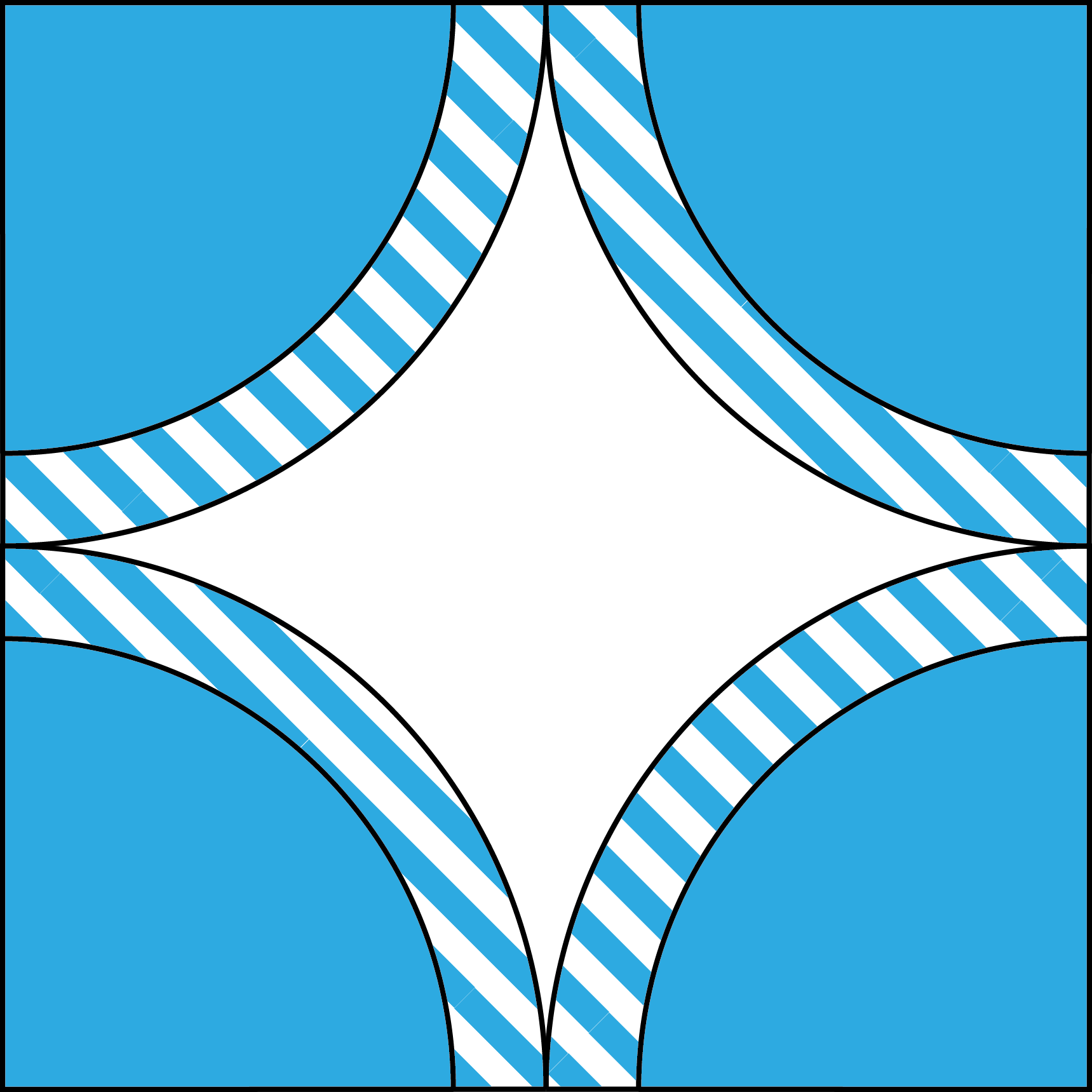}
\end{minipage} \\
\begin{minipage}[t]{3mm} 
\large b) 
\end{minipage}
\begin{minipage}[t]{60mm}
\vspace{-10pt}
\includegraphics[width=\textwidth]{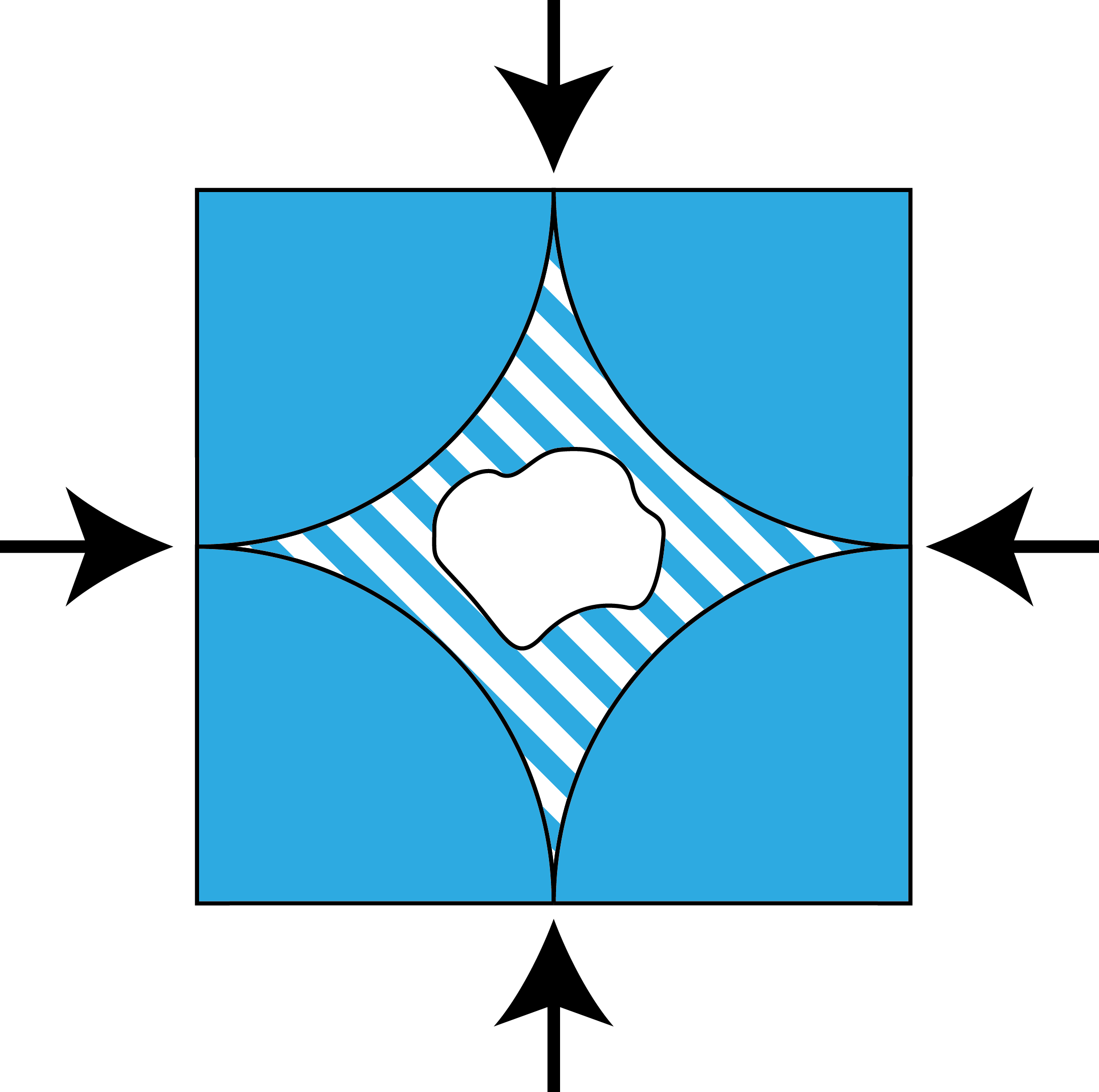}
\end{minipage} \\ 
\caption{\label{fig:const-porosity} End-member model for constant porosity. The illustration shows melting of a lattice sphere packing. a) Melting reduces the radius of the spheres. b) Snow compacts and ice grains stay in contact to each other. The density of the snowpack increases, but the porosity remains constant.}
\end{figure}

The other end-member representing a lower limit for the porosity change during snow melt is based on the assumption that the porosity remains constant due to ongoing compaction of the snowpack. 
This assumption can be illustrated with a packing of spheres where snow melt reduces the radii of the spheres and the snowpack compacts until the spheres are again in contact with each other. 
This end-member is illustrated in Figure \ref{fig:const-porosity} for a cubic lattice packing of spheres.

The constant porosity of this end-member model also leads to constant properties of the porous frame. In this model, in terms of wave propagation, only the pore fluid properties change with increasing water saturation.
While the porosity remains constant over the process of snow melt, the total volume of the snowpack reduces due to snow compaction. 
If the snowpack is assumed to compact only in vertical direction, the change of thickness of the snow cover $\Delta h$ follows the relationship
\begin{equation}
\Delta h = h_0 \frac{\Delta \rho}{\rho_0},
\end{equation}
where $\rho_0$ is the initial density and $\Delta \rho$ is the change in density.
The relation between density and liquid water saturation can be obtained using equation (\ref{eq:density}).
For example, a dry snowpack with a density of 400~kg/m$^3$ melting with constant porosity leads to a reduction of the snowpack hight of $\sim$14~\% when reaching a saturation of $S_w = 0.1$.
The use of this end-member as an upper bound of snow compaction is based on the consideration that a stronger increase in density is only possible if the geometrical constellation of the packing is changed. A stronger compaction is either possible if the original packing was not a dense packing and is changing the constellation toward a denser packing or if the spheres develop a dipolar distribution of radii.

For the same amount of melted snow, the pore water saturation increases faster for the constant porosity end-member than for the constant density end-member. 
The percentage of the ice frame that has melted can be estimated as
\begin{equation}
\% \, melt \cong  \frac{ \phi S_{\rm  \textsc{w}} \rho_{\rm  \textsc{w}} }{ (1-\phi) \rho_{\rm  \textsc{i}} + \phi S_{\rm  \textsc{w}} \rho_{\rm  \textsc{w}}} ,
\end{equation}
where the porosity $\phi$ is a function of water saturation $\phi (S_{\rm  \textsc{w}})$ for the end-member with constant density. 
Starting with a dry snowpack with a density of 400~kg/m$^3$ ($\phi \cong 0.56$) that melts without any pore water added or removed, a pore water saturation of $S_{\rm \textsc{w}} = 0.1$ corresponds to 12 \% melt of the ice mass for the end-member with constant porosity and 16 \% for the end-member with constant density. 

\section{Results}

\subsection{Change of velocity with increasing water content}

In this section phase velocities and attenuation obtained with plane wave solutions of Biot's differential equations for wave propagation in porous materials are shown for the presented model.
The model has free degrees in pore water saturation, frame porosity, and frequency.
These model parameters are inspected for an acoustic experiment to estimate liquid water content in snow and with consideration of the two end-member models presented above.

\begin{figure}
\centering
\begin{minipage}[t]{3mm}
\large a) 
\end{minipage}
\begin{minipage}[t]{80mm}
\vspace{-10pt}
\includegraphics[width=1\textwidth]{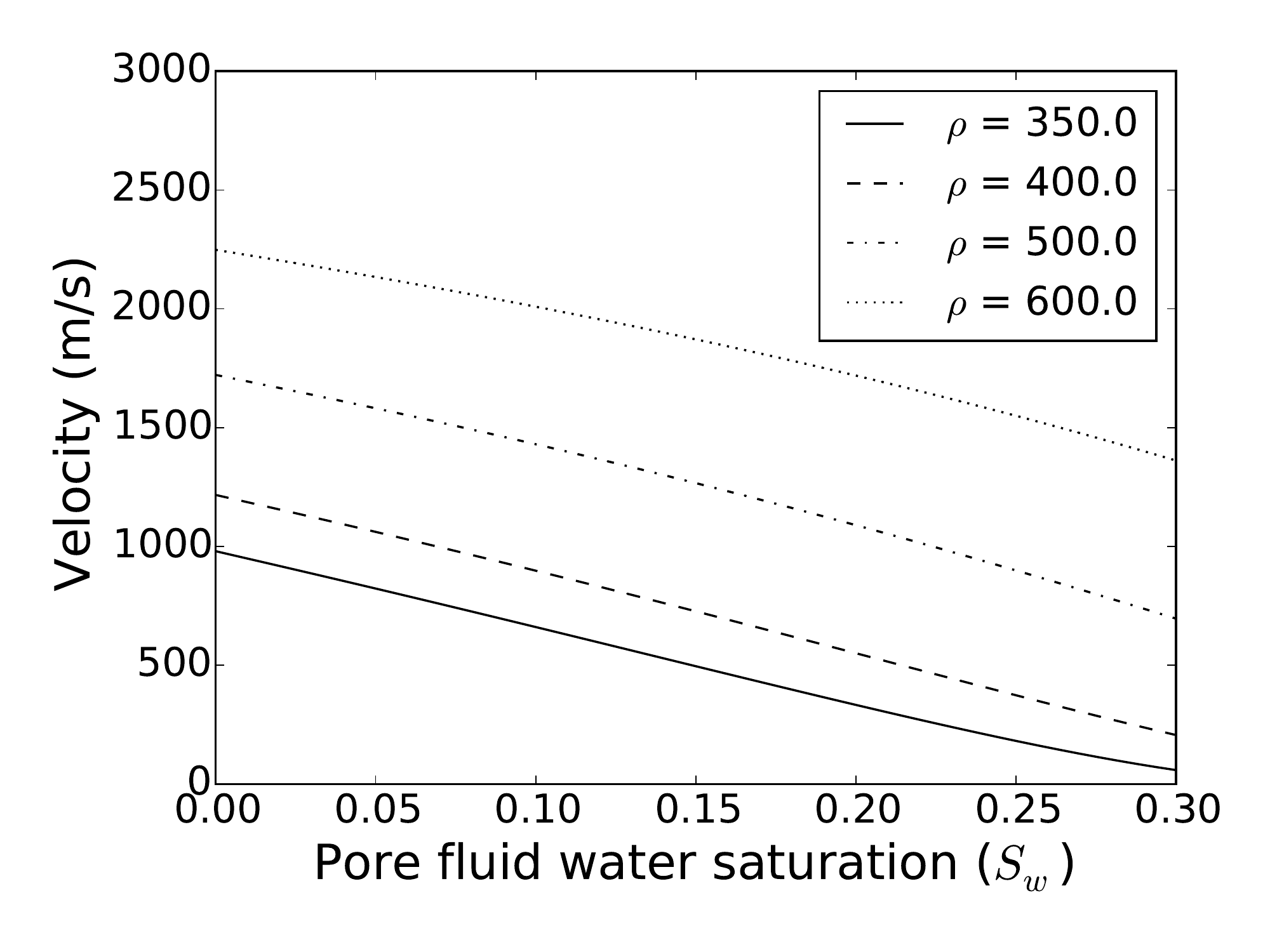}
\end{minipage} \\
\begin{minipage}[t]{3mm}
\large b) 
\end{minipage}
\begin{minipage}[t]{80mm}
\vspace{-10pt}
\includegraphics[width=1\textwidth]{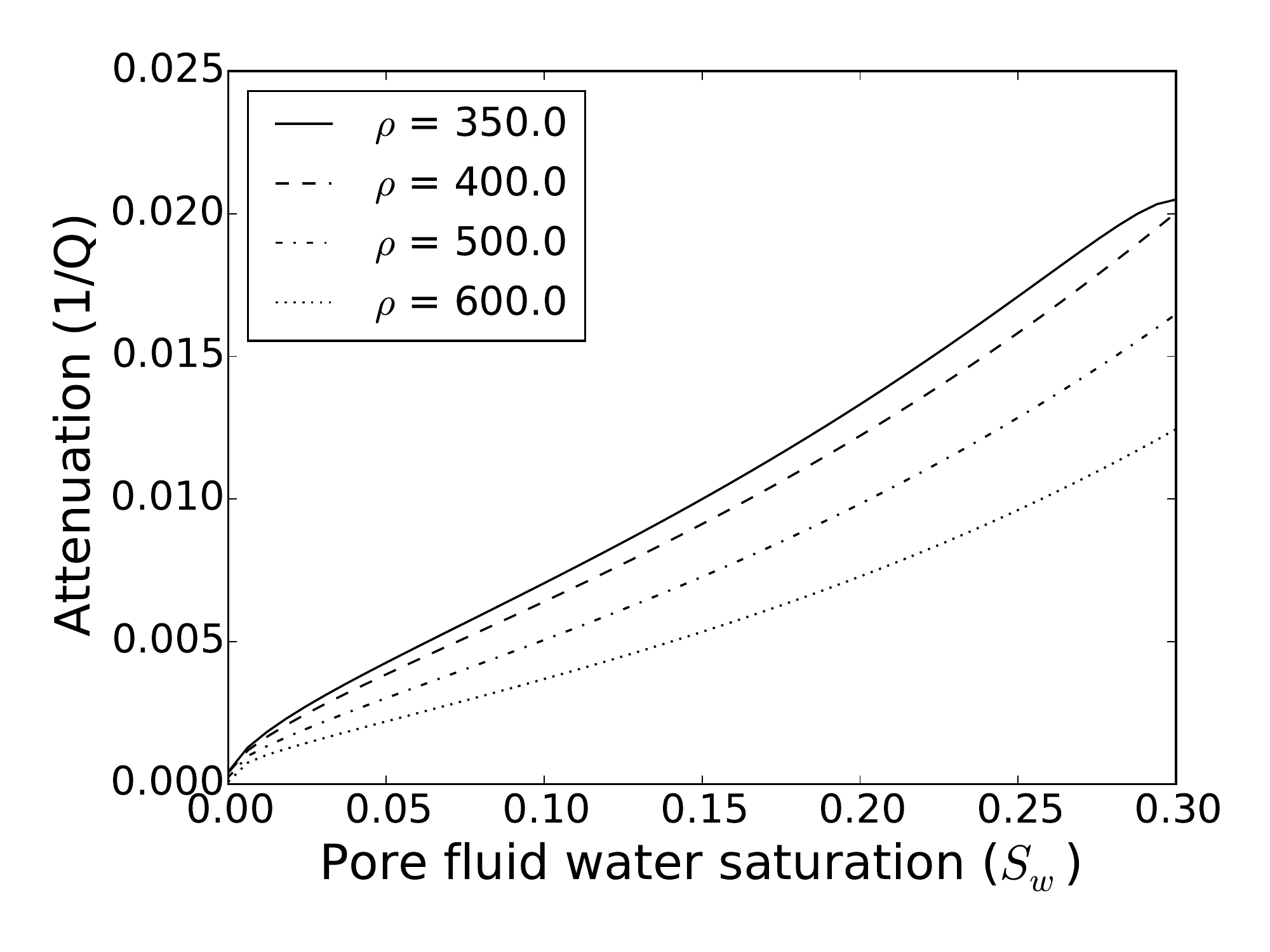}
\end{minipage} \\ 
\caption{\label{fig:vel4sat} Change of a) phase velocity of the first compressional wave at 1~kHz with increasing liquid water saturation in the pore space and b) the corresponding increase of attenuation.}
\end{figure}

A general trend of decreasing compressional wave velocity with increasing liquid water content in the pore space can be observed. 
Figure \ref{fig:vel4sat}a) shows the phase velocity variations at 1 kHz for the first compressional wave for snowpacks with different densities.
Higher velocities are observed for denser snowpacks because of the stiffer ice frame for snow with lower porosity.
Even though the trend is not exactly linear, the decrease of phase velocity with increasing water saturation is roughly 300~m/s less for an increase of 0.1 in liquid water saturation.
This is a significant change in velocity and it should be straight forward to detect such velocity changes in field measurements. 

\begin{figure}
\centering
\includegraphics[width=80mm]{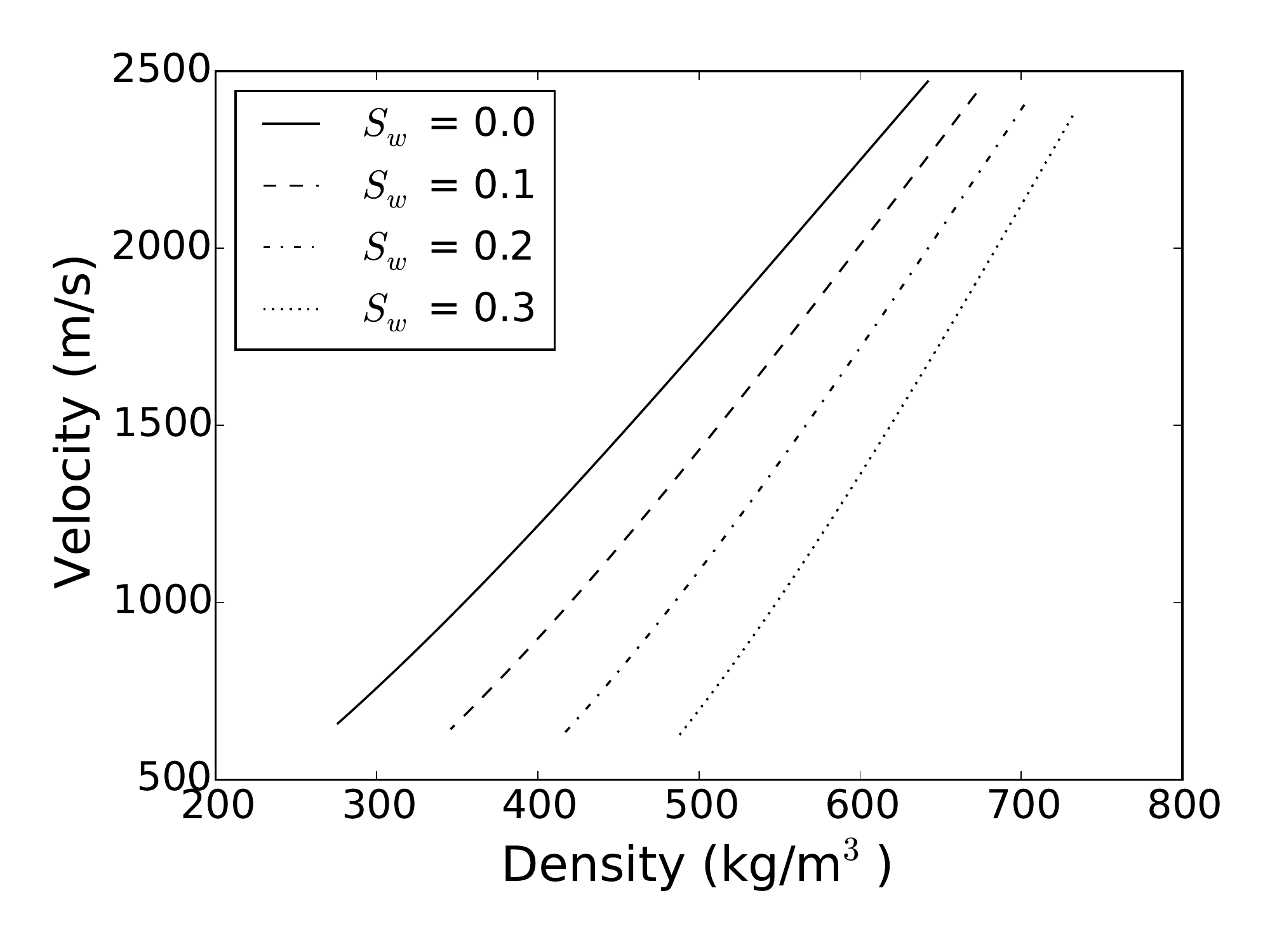}
\caption{\label{fig:known-sat} 
Phase velocity of the first compressional wave at 1~kHz as a function of density for different water saturations.}
\end{figure}

The phase velocity of the first compressional wave not only changes with pore water saturation but also with the density of the entire snowpack. This change in velocity is due to the stiffer ice frame in high density, low porosity snow.
The increase of phase velocity is not exactly linear and varies with liquid water saturation as is shown in Figure \ref{fig:known-sat}, but is roughly an increase of 500 m/s for an increase of 100 kg/m$^3$ in density.
The strong gradient allows for a robust estimate of snow porosity in an acoustic experiment to determine liquid water saturation in combination with an independent density measurement.

The attenuation for the first compressional wave is shown in Figure \ref{fig:vel4sat}b). 
The attenuation increases with increasing water content. For lighter snow the attenuation increases stronger than for denser snow as the stiffness of the equivalent fluid is proportionally larger compared to the lower stiffness of the ice frame.

\begin{figure}
\centering
\begin{minipage}[t]{3mm}
\large a) 
\end{minipage}
\begin{minipage}[t]{80mm}
\vspace{-10pt}
\includegraphics[width=1\textwidth]{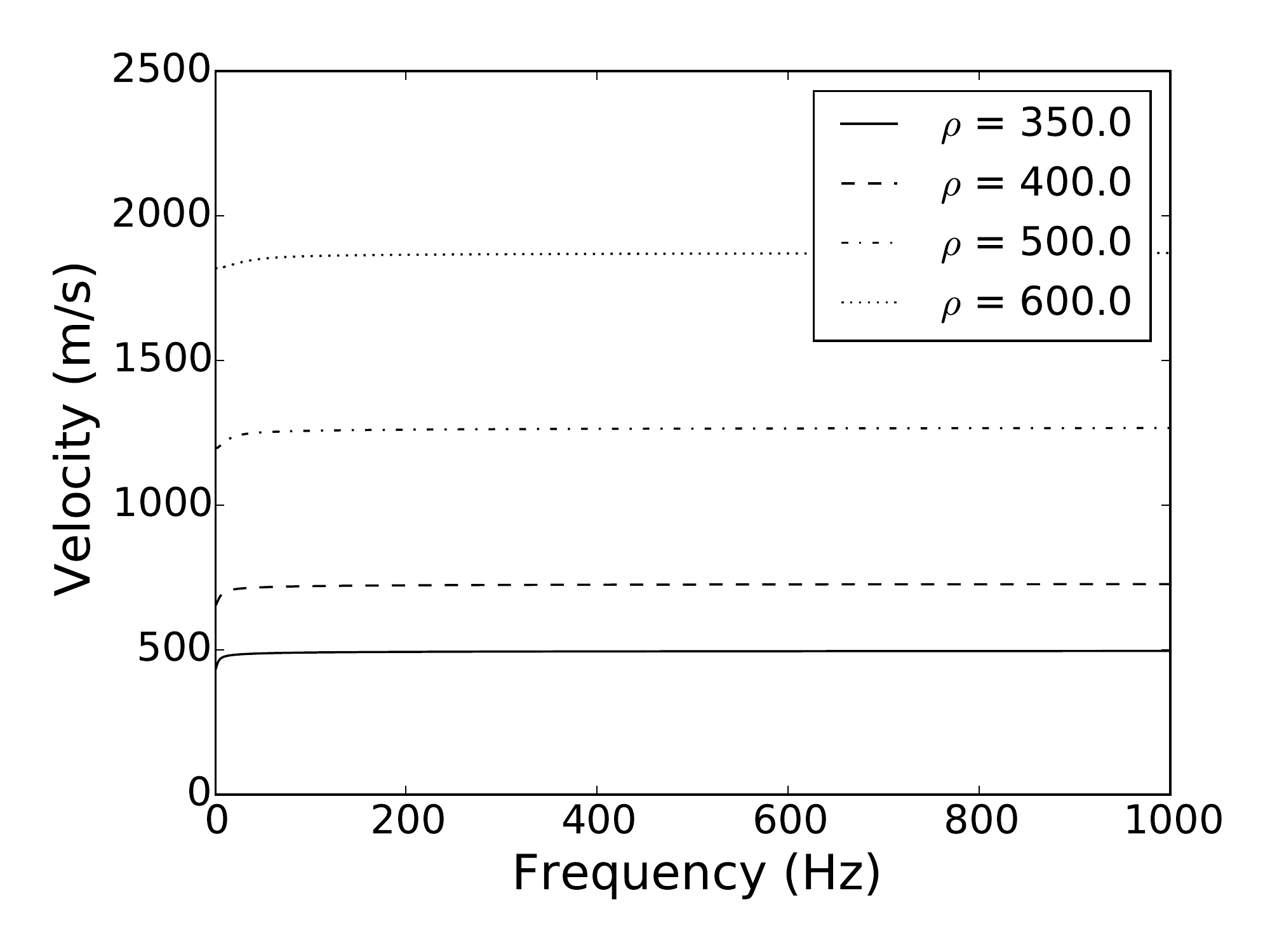}
\end{minipage} \\
\begin{minipage}[t]{3mm}
\large b) 
\end{minipage}
\begin{minipage}[t]{80mm}
\vspace{-10pt}
\includegraphics[width=1\textwidth]{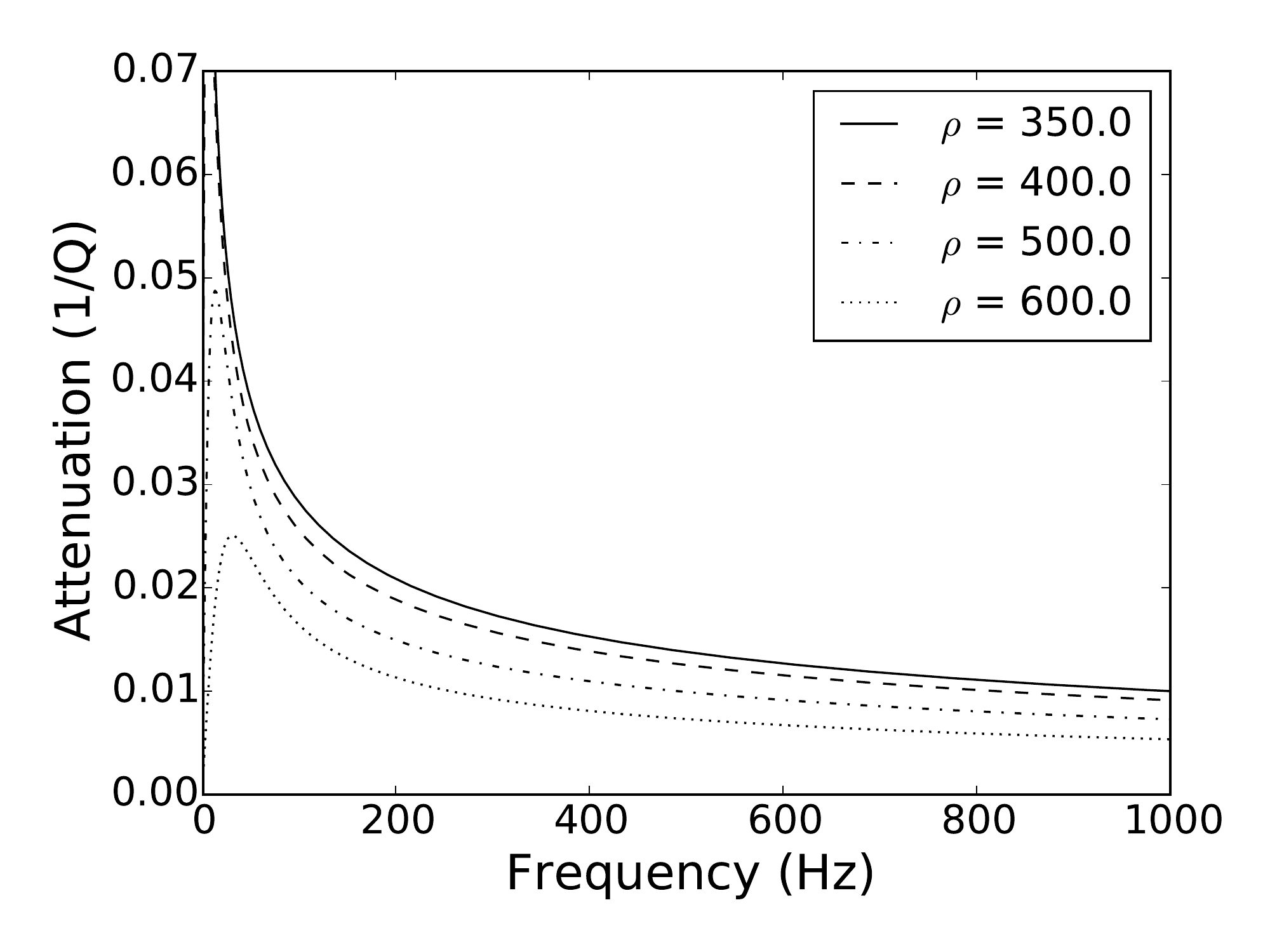}
\end{minipage} \\ 
\caption{\label{fig:freqdep} Predicted a) phase velocity and b) attenuation for the first compressional wave for a liquid water saturation of $S_w = 0.15$ in the pore space.}
\end{figure}

Figure \ref{fig:freqdep} shows the frequency dependence of the phase velocity and attenuation shown in Figure \ref{fig:vel4sat}. The values for a saturation of 0.15 and a frequency of 1~kHz correspond to each other in these two Figures. 
It can be seen that there is only little dispersion in the phase velocities that is concentrated around very low frequencies.
Also the attenuation is highest at low frequencies and exponentially decays for higher frequencies.
Again, the attenuation is highest for lighter snow and and less pronounced for denser snow due to the stiffness proportions of  the frame and the equivalent fluid.

\subsection{Requirement of an independent snow density measurement}

The two end-member models are used to asses the necessity of an independent density measurement to estimate liquid water content in snow with acoustic waves.
An independent measurement of snow density is a limitation for estimating liquid water content.
Automatic measurements of snow density are cumbersome and therefore a density measurement mostly includes manual work, also prohibiting self acting liquid water content measurements.
In addition, snowpacks can exhibit strong lateral variations in snow density. Especially in the presence of liquid water in the snowpack. An independent density measurement does not necessarily sample the same space that the remote sensing method does, thereby introducing an additional uncertainty.
If the two end-member models show the same variation with liquid pore water content, such a measurement opens up the possibility to estimate snow density and liquid water content with a single acoustic experiment.



\begin{figure}
\centering
\includegraphics[width=80mm]{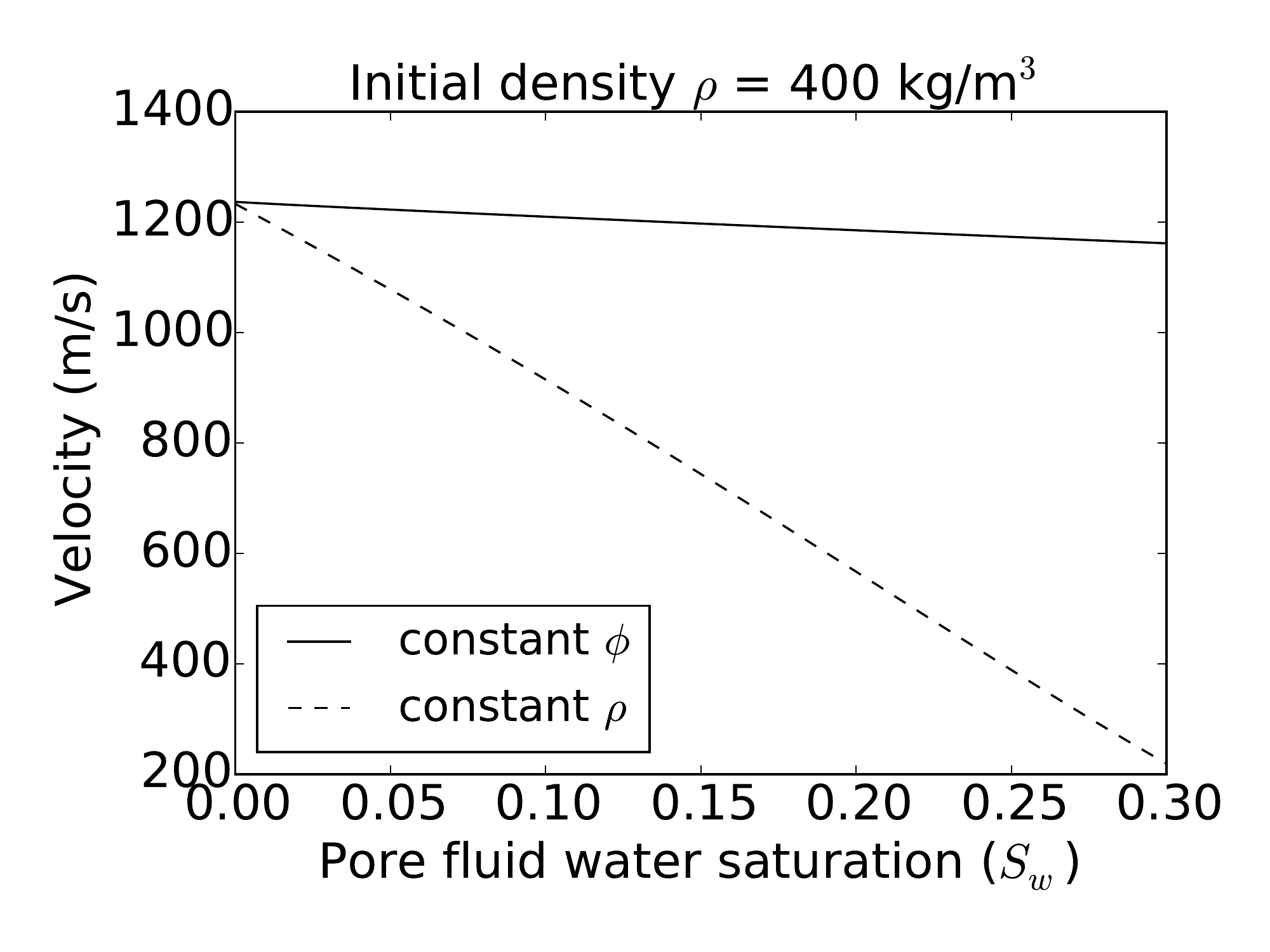}
\caption{\label{fig:sat-vel} 
Phase velocity of the first compressional wave for a snowpack with initial density of 400 kg/m$^3$ as a function of pore water saturation for the end-member model of (solid line) a constant porosity and (dashed line) constant density.}
\end{figure}

Figure \ref{fig:sat-vel} shows the decrease of compressional wave velocities as a function of pore water saturation $S_{\rm w}$ for both end-member models.
To compare the two end-members the curves are shown for a dry density of 400 kg/m$^3$ and the density of the constant porosity model is adjusted with increasing liquid water content according to equations (\ref{eq:density}) and (\ref{eq:porchange}).
The two curves strongly spread, indicating that a combination with an independent density measurement is important to estimate the liquid water content in snow when measuring the phase velocity of the first compressional wave only.

\begin{figure}
\centering
\includegraphics[width=80mm]{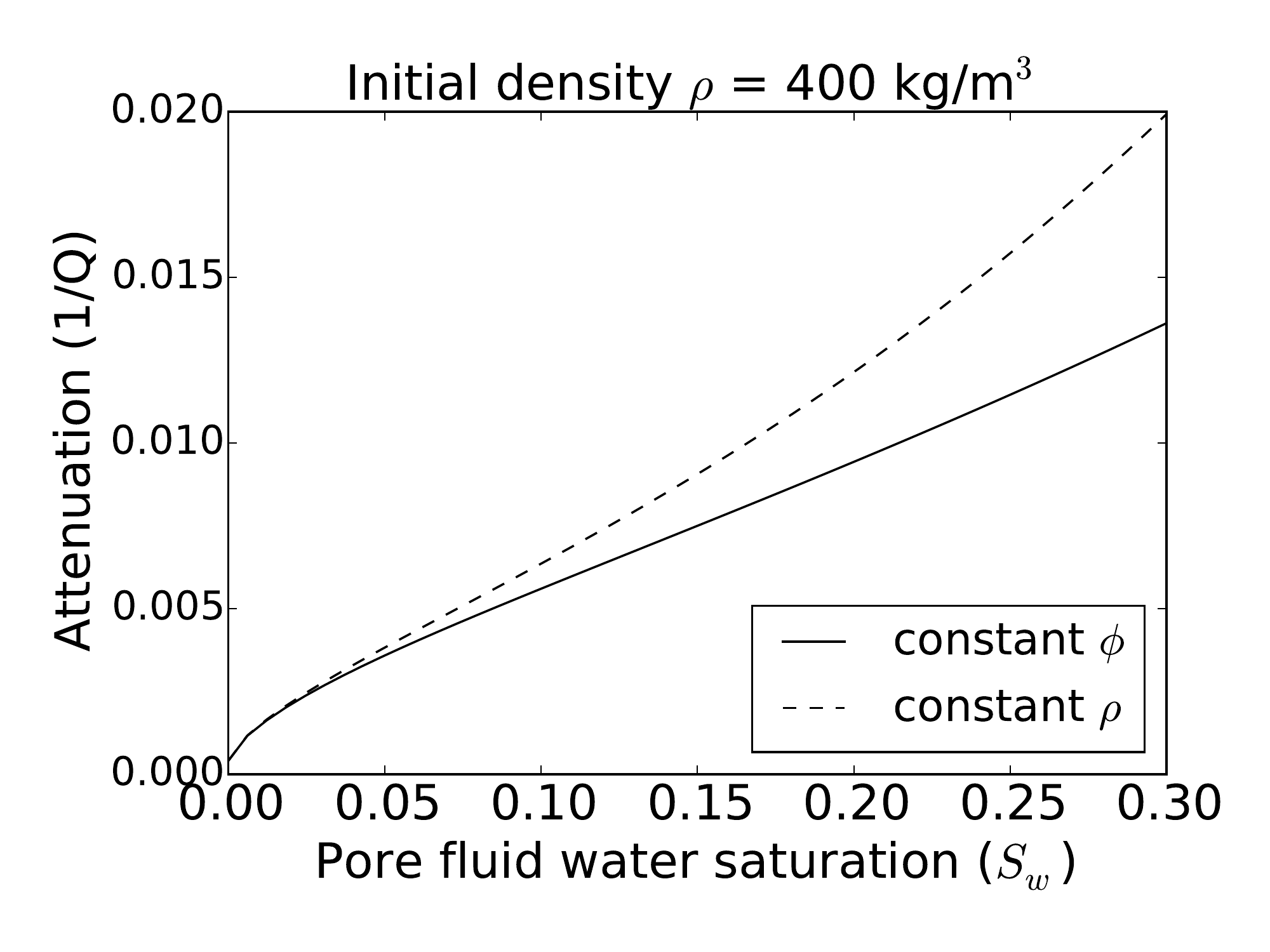}
\caption{\label{fig:sat-att} 
Attenuation of the first compressional wave for a snowpack with initial density of 400 kg/m$^3$ as a function of pore water saturation for the end-member models of (solid line) a constant porosity and (dashed line) constant density.}
\end{figure}

The plane wave attenuation for compressional waves of the first kind are shown in Figure \ref{fig:sat-att}. Again the two curves for the two end-member models are shown to indicate minimum and maximum levels to be expected for a dry snowpack density of 400 kg/m$^3$.
Especially for small amounts of liquid water in the pore space the two curves are close together. Indicating that the attenuation of the first compressional wave could be used to assess density and liquid water content in combination with the phase velocity.
However, this is possible only if the geometrical effects leading to a decreasing signal amplitude are well known and additional factors leading to attenuation as, for example, scattering of the signal from lateral heterogeneities can be excluded or reconstructed.

\begin{figure}
\centering
\includegraphics[width=0.6\textwidth]{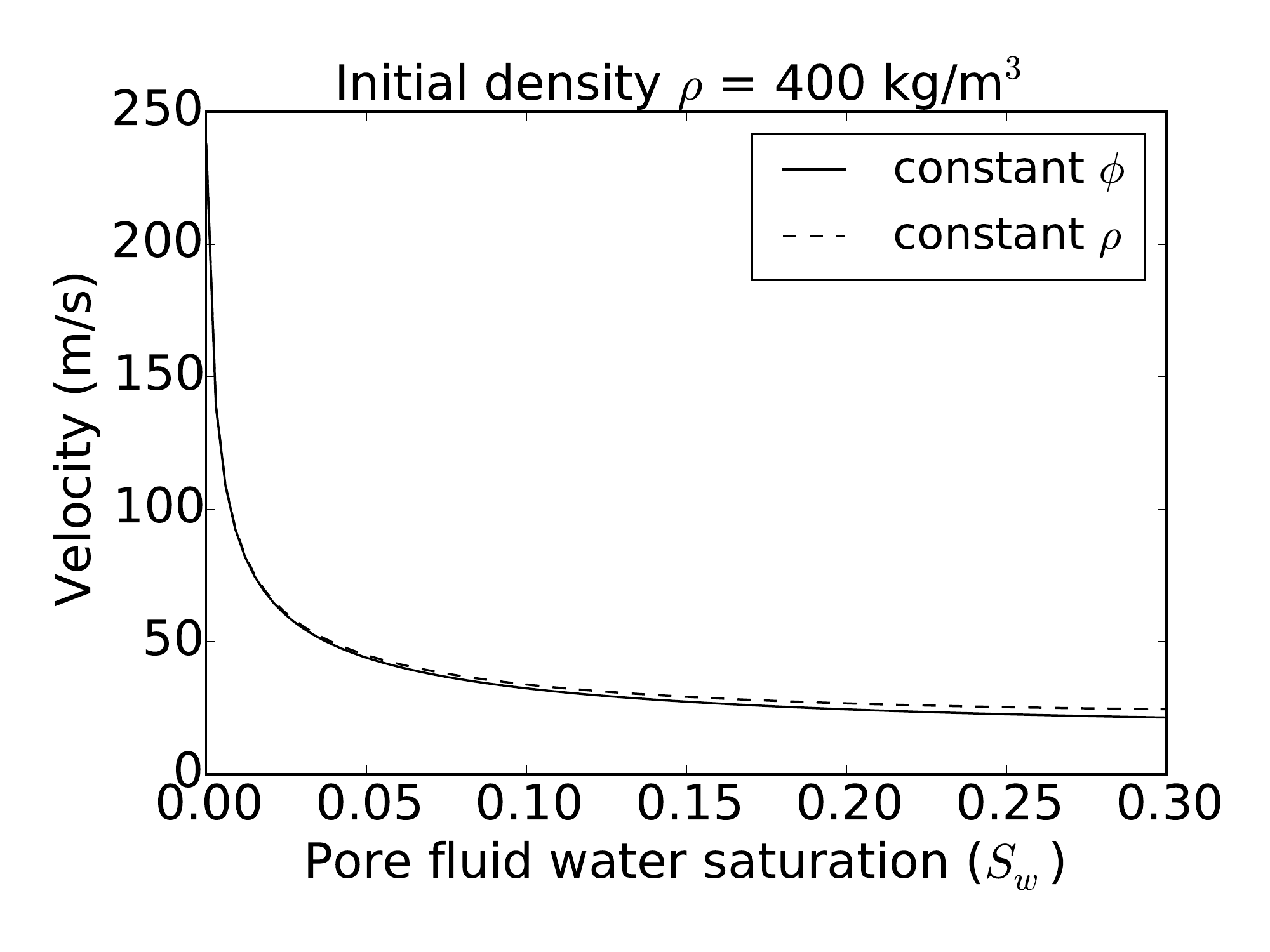}
\caption{\label{fig:sat-V2} 
Phase velocity for the second compressional wave in a snowpack with a dry density of 400 kg/m$^3$ as a function of pore water saturation. The solid line corresponds to the end-member of constant porosity and the dashed line to the end-member of constant density.}
\end{figure}

Theory for poroelastic materials predicts a second compressional wave mode where the particle velocity of the pore fluid is out of phase with the particle motion of the solid frame. This wave mode, sometimes called the 'slow' wave is difficult to observe in geological formations and is generally seen only in laboratory experiments \citep{jocker:2009}.
Yet, the properties of snow allow, at least under some conditions, the propagation of this second compressional wave mode and the `slow' wave has been measured in field conditions \citep{oura:1952,ishida:1965,johnson:1982}. 
The second compressional wave is of interest because it is sensitive mostly to the properties of the pore space. 
It is therefore not surprising that in Figure \ref{fig:sat-V2} the phase velocity for the second compressional wave at 1~kHz shows almost the same variation with increasing pore water saturation for both snow compaction end-member models.

\begin{figure}
\centering
\includegraphics[width=0.6\textwidth]{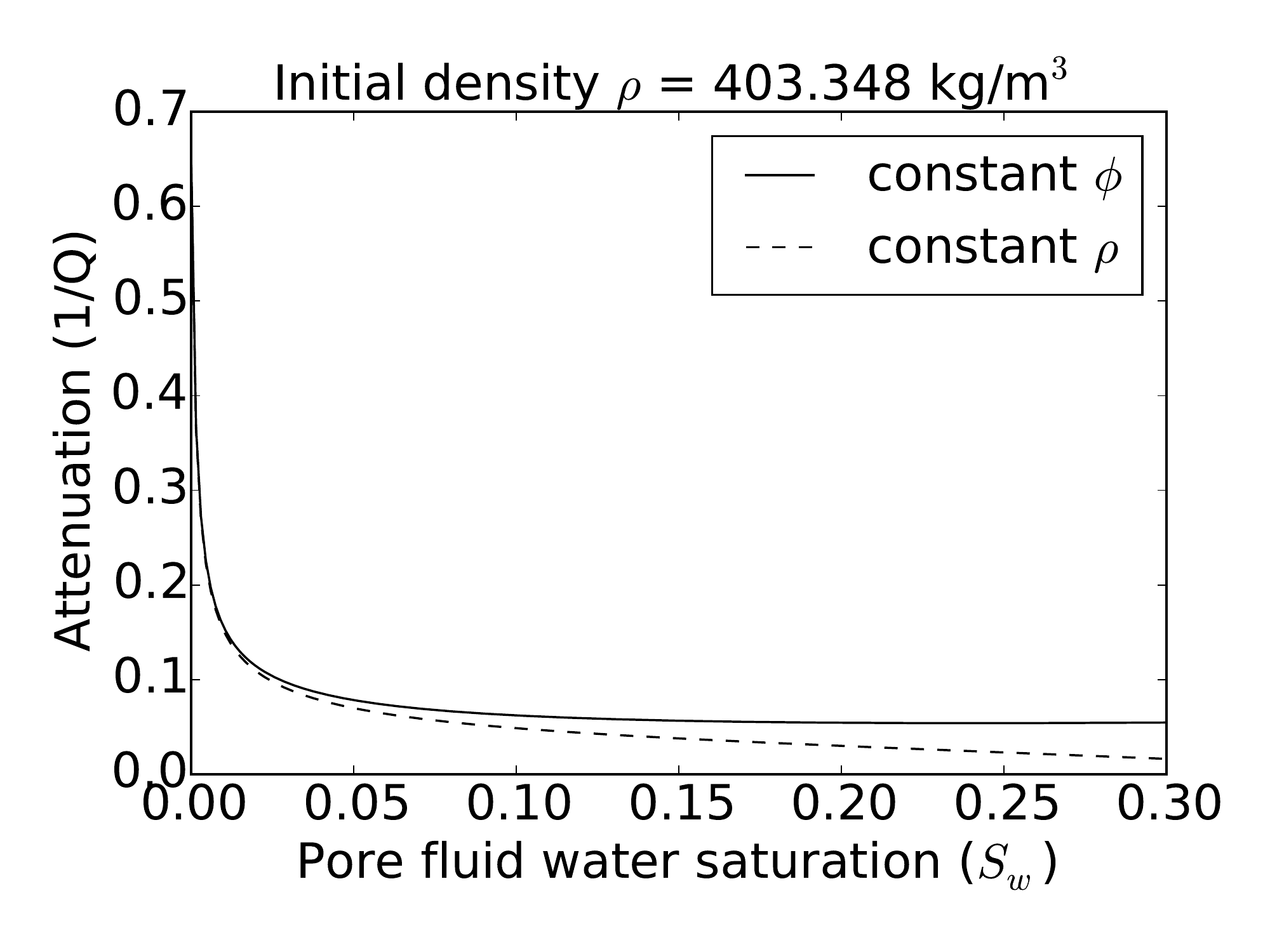}
\caption{\label{fig:sat-Q2} 
Attenuation for the second compressional wave in a snowpack with a dry density of 400 kg/m$^3$ as a function of pore water saturation. The solid line corresponds to the end-member of constant porosity and the dashed line to the end-member of constant density.}
\end{figure}

Also the attenuation of the second compressional wave that is shown in Figure \ref{fig:sat-Q2} has only little difference between the two end-member models.
Interestingly the attenuation of the second compressional wave reduces with increasing liquid water content making the detection of this wave mode more prospecting. 
A potential limitation for the detection of the second compressional wave in wet snow is the lower porosity and accompanying higher stiffness of the frame that increases the attenuation of the slow wave. 
No observations of the second compressional wave are reported for densities larger than 400 kg/m$^3$ \citep{johnson:1982,sommerfeld:1982}.

Due to geometrical effects and a number of mechanisms that can influence attenuation measurements, much more elaborate techniques are necessary to properly evaluate attenuation measurements compared to the measurement of a wave velocity. 
A measurement of the first and second compressional wave velocity would therefor be the preferred way to estimate the density and liquid water content with a single acoustic experiment, supposed that the second compressional wave can be measured in wet snow.
Numerical simulations show evidence that this is indeed the case under some conditions.

\section{Field example}

\begin{figure}
\centering{
\includegraphics[width=80mm]{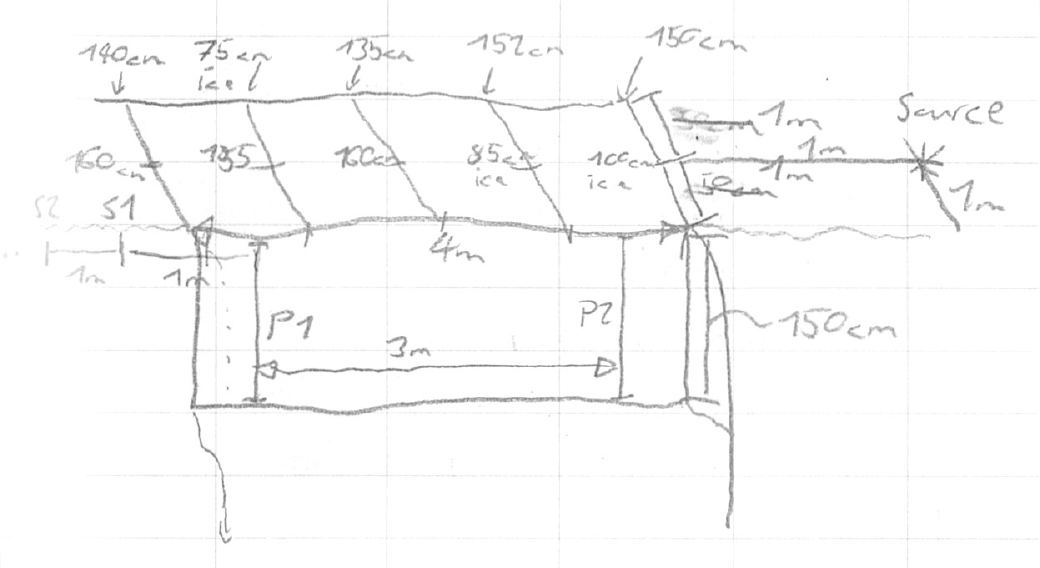}}
\caption{\label{fig:meas-setup} Sketch of the acoustic experiment showing the locations of the acoustic source at the snow surface and the accelerometers in the two snow pits.
}
\end{figure}

\begin{figure}
\centering{
\includegraphics[width=80mm]{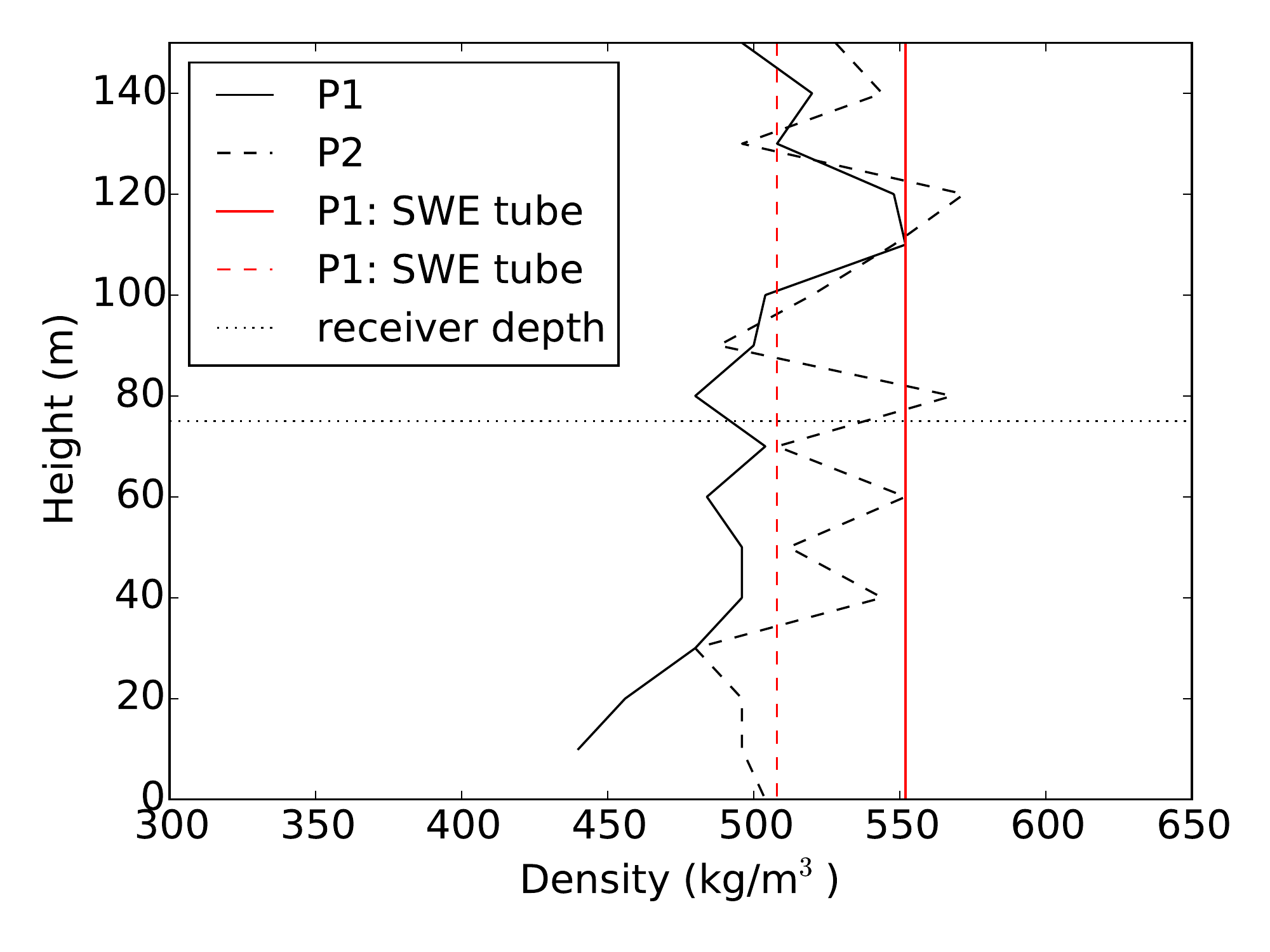}}
\caption{\label{fig:density} Density profiles in the two snow pits P1 and P2. The density was measured every 10~cm and for the whole snowpack with a snow water equivalent (SWE) tube.
}
\end{figure}

In this section a simple acoustic experiment in snow is used to illustrate the possibility to measure liquid water content in snow with acoustic waves.
A sketch of the geometrical configuration of the experiment is shown in Figure \ref{fig:meas-setup}. A rock was placed at the snow surface that would become an acoustic source later in the experiment when hit with a sledge hammer. Two snow pits were excavated 1 m and 3m from the location of the acoustic source and an accelerometer was placed 0.75 m below the snow surface in each snow pit.
Density profiles for each snow pit were recorded by measuring the weight of a container with a volume of 250~cm$^3$ that was filled by pressing it horizontally into the snow at 0.1 m intervals on a vertical profile. In parallel, a cylindrical tube with a cross-section of 30 cm$^{2}$ was vertically pressed down to the bottom of the snow pack allowing to recover and weight a vertical snow column.
The resulting density profiles and the average density measurements with the snow tube are shown in Figure \ref{fig:density}.

\begin{figure}
\centering
\begin{minipage}[t]{3mm}
\large a) 
\end{minipage}
\begin{minipage}[t]{80mm}
\vspace{-10pt}
\includegraphics[width=\textwidth]{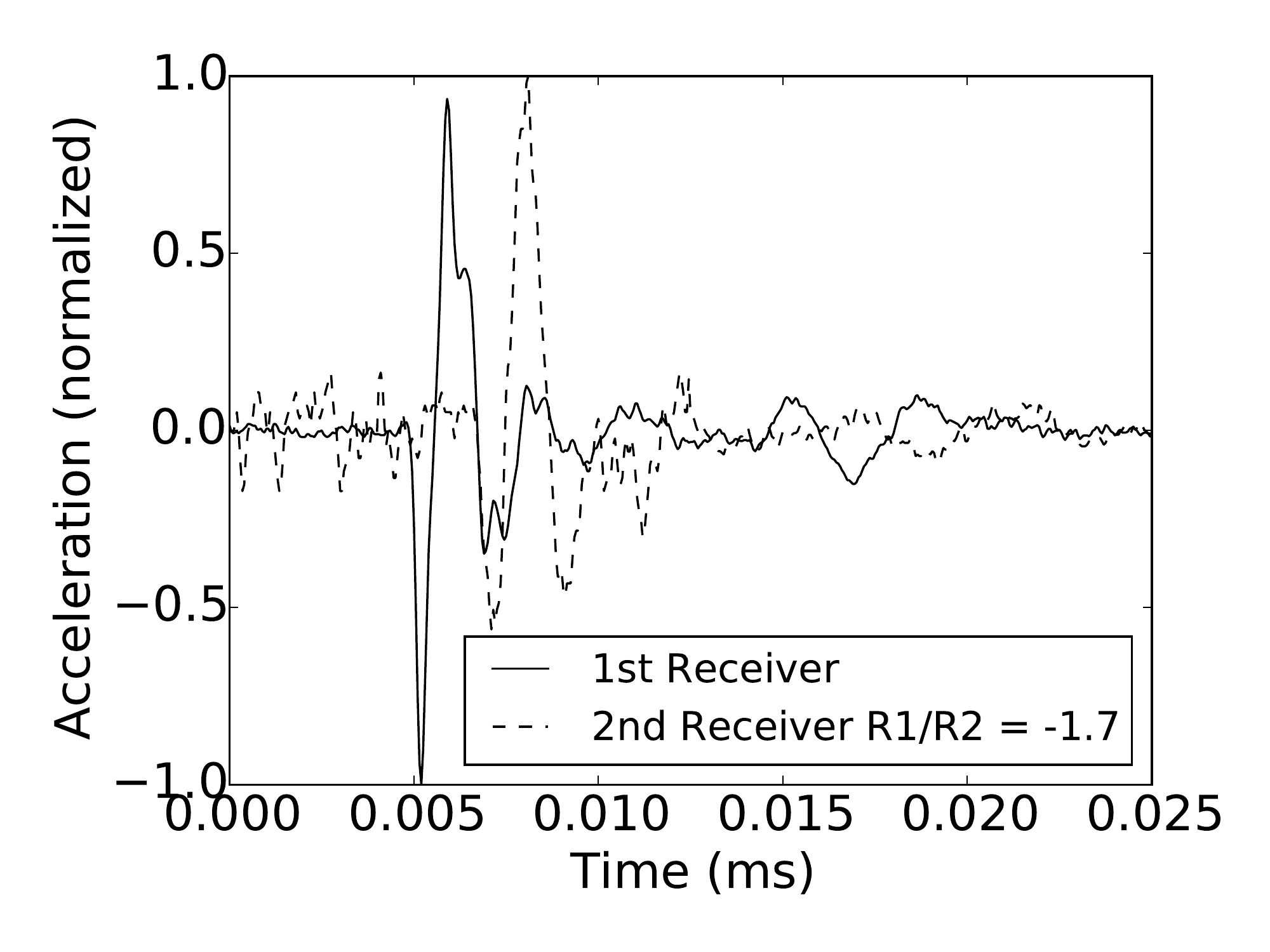}
\end{minipage} \\
\begin{minipage}[t]{3mm} 
\large b) 
\end{minipage}
\begin{minipage}[t]{80mm}
\vspace{-10pt}
\includegraphics[width=\textwidth]{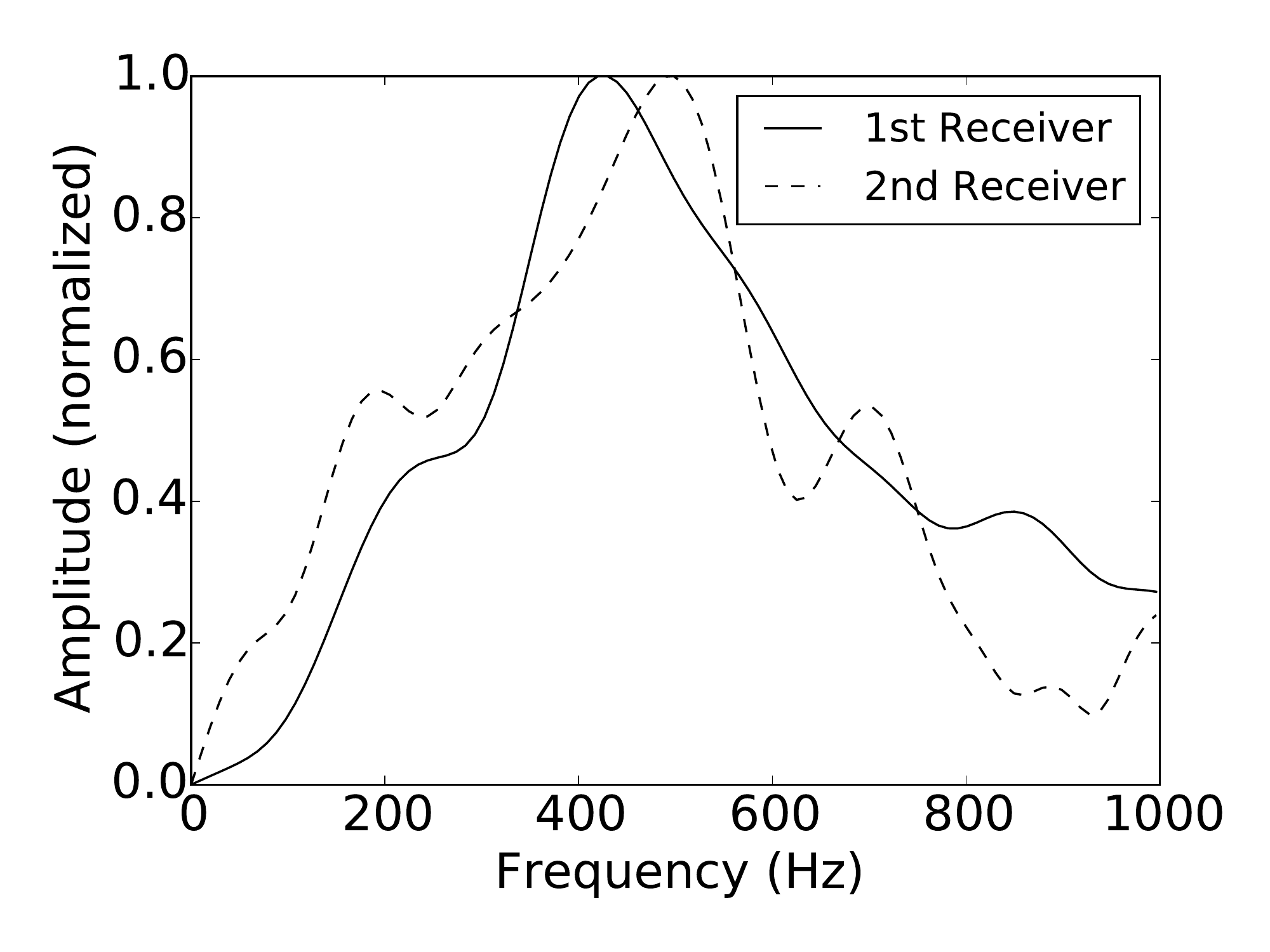}
\end{minipage}
\caption{\label{fig:recordings} Normalized a) acceleration and b) corresponding amplitude spectra for the measurements in the two snow pits. The acceleration of the second receiver is corrected for spherical spreading and multiplied by 1.7 to have the same maximal amplitude as the first receiver.
}
\end{figure}

\begin{figure}
\centering
\begin{minipage}[t]{3mm}
\large a) 
\end{minipage}
\begin{minipage}[t]{80mm}
\vspace{-10pt}
\includegraphics[width=\textwidth]{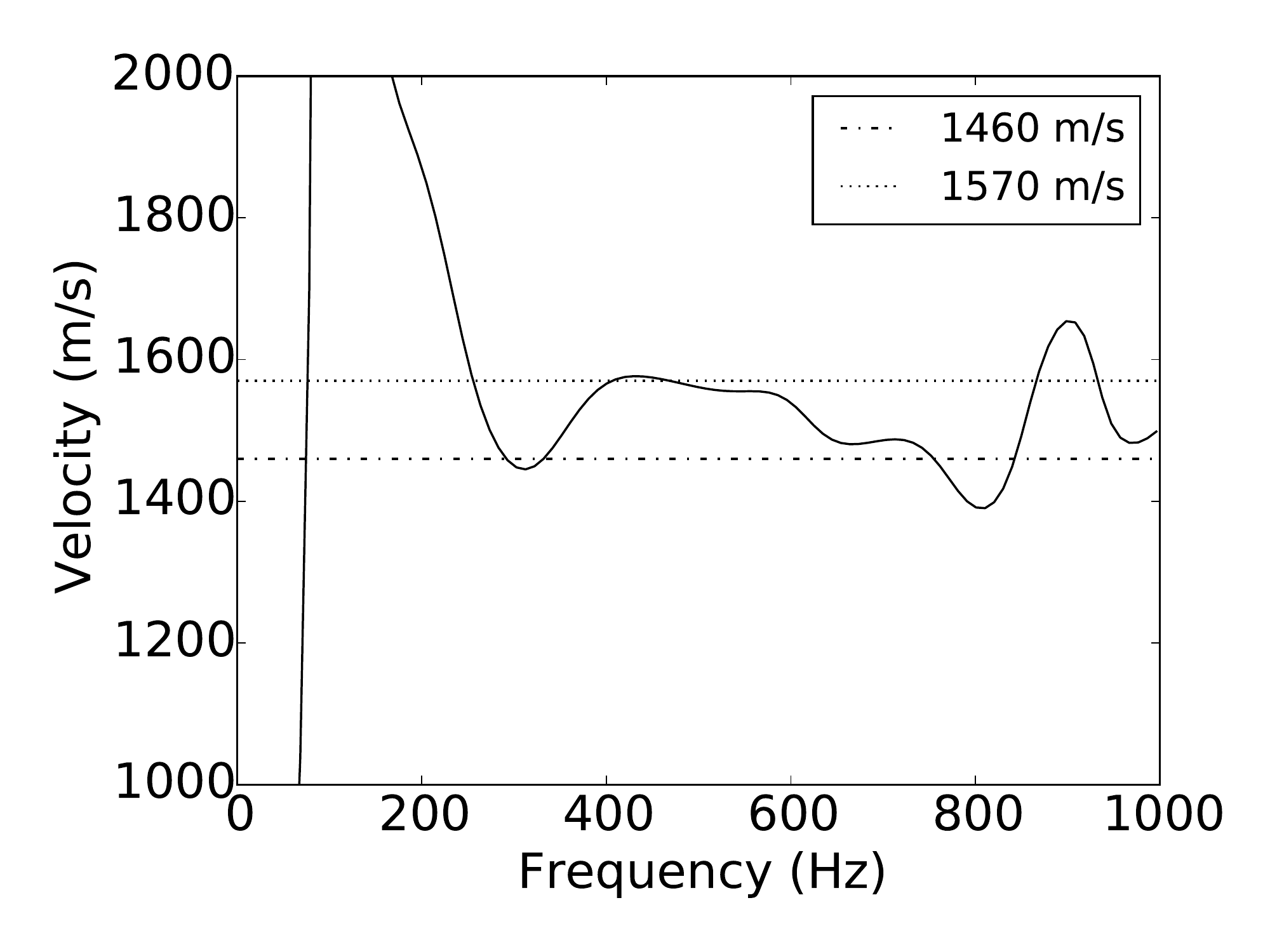}
\end{minipage} \\
\begin{minipage}[t]{3mm} 
\large b) 
\end{minipage}
\begin{minipage}[t]{80mm}
\vspace{-10pt}
\includegraphics[width=\textwidth]{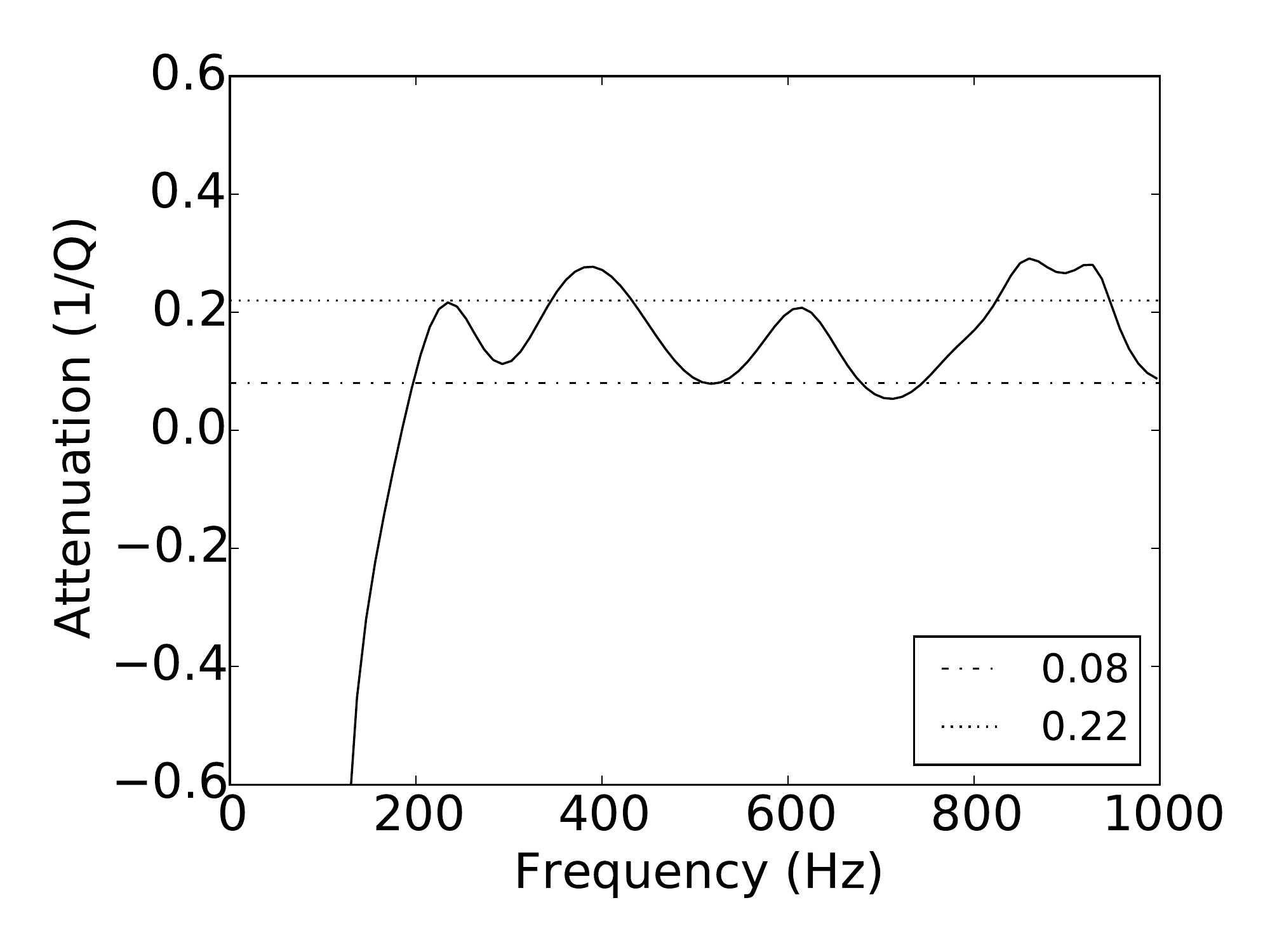}
\end{minipage}
\caption{\label{fig:estimates} Estimated a) phase velocity and b) attenuation from the measurements shown in Figure \ref{fig:recordings}. The amplitudes were corrected for spherical geometrical spreading to compute the attenuation. The minimum and maximum values evaluated in Table \ref{tab:results} are indicated with horizontal lines.
}
\end{figure}

For the experiment, the recording of the accelerometers was started with a sampling rate of 20 kHz and the acoustic source was excited.
Figure \ref{fig:recordings}a shows the section of the recordings where the acoustic signal reached the accelerometers. The absolute zero timing was randomly chosen in front of the arrival of the signal at the first accelerometer. 
The amplitude spectra of the two recorded signals are shown in Figure \ref{fig:recordings}b). The main frequency contribution of the signal is between 200 Hz and 800 Hz.
To obtain the phase velocity and the frequency dependent attenuation the two signals were Fourier analyzed.
The phase spectrum was used to compute the phase velocity and the amplitude spectrum was used to compute the attenuation \citep{sachse:1978,oconnel:1978,pialucha:1989,reine:2009}. 
The corresponding results are shown in Figure \ref{fig:estimates}a) and \ref{fig:estimates}b), respectively.

As there is little to no dispersion expected in this experiment (see Figure \ref{fig:freqdep}a)), the phase velocity is best evaluated at the central frequency of the propagating pulse.
Which is around 500~Hz for this experiment. 
The phase velocity at other frequencies are used to obtain an estimate on the precision of the velocity measurement.
Based on Figure \ref{fig:estimates}a) the maximum and minimum velocity for wave propagation between the two sensors was chosen to be 1460 m/s and 1570 m/s, respectively.

To obtain an estimate for the volume fraction of liquid water content the phase velocity and the independent density measurement were inverted for porosity $\phi$ and liquid water saturation of the pore space $S_w$ as follows:
\begin{enumerate}
\item Use a least square cost function equally weighting errors in velocity and density and chose random initial values for the unknown porosity and liquid water saturation.
\item Calculate the properties of a Biot-type porous material and an equivalent pore fluid for the chosen initial values using Table \ref{tab:snowmodel} and \ref{tab:porefluid} and equations (\ref{eq:effvisc}), (\ref{eq:effmod}), and (\ref{eq:effdens}).
\item Compute the phase velocity for the resulting material for the plane wave solution of Biot's equation for wave propagation in porous materials (\ref{apx:planewave}).
\item Evaluate the cost function and optimize the initial values by repeating steps 2 to 4 for adjacent values of the initial guess according to the Nelder-Mead scheme \citep{nelder:1965}.
\end{enumerate}

The resulting porosity and liquid water saturation as well as the corresponding volume fraction of liquid water content are shown in Table \ref{tab:results}.
Depending on the velocity and density measurements the liquid water content for the examined snowpack is estimated to be between 4 \% and 8 \% and can be classified as wet snow (3) on a scale from (1) which corresponds to 0 \% Vol to (5) which corresponds to more than 15 \% Vol \citep{fierz:2009}.
This corresponds with the observations that liquid water was discharging from the snowpack, but water could not be pressed out of the snow by squeezing the snow in by hands. 
No alternative quantitative measurements of liquid water content have been made and the rule of thumb observations are an admittedly weak source for comparison.
However, it is interesting to see that the estimate for the porosity is relatively constant in the range of the estimated precision of the measurements and that the liquid water content depends more strongly on the precision of the density measurement than on the phase velocity estimate.

In future research the liquid water estimates with acoustic methods will be compared to quantitative measurements with alternative methods.
Also the spatial sensitivity of the measurement and the strongly heterogeneous distribution of the liquid water in the pore space will be investigated.

\begin{table}
\caption{Estimated snow porosity and liquid water content for minimum and maximum estimated phase velocities and densities.}
\begin{center}
\begin{tabular}{lll}
\hline
 					&	$V_{est}$ = 1460			&	$V_{est}$ = 1570 \\
\hline
$\rho$ = 408 kg/m$^3$	&	$\phi$ = 0.50 $S_w$ = 0.10 	&	$\phi$ = 0.48 $S_w$ = 0.07	\\
					&	 5 \% Vol					&	4 \% Vol					\\
\hline
$\rho$ = 552 kg/m$^3$	&	$\phi$ = 0.50 $S_w$ = 0.18 	&	$\phi$ = 0.48 $S_w$ = 0.15	\\
					&	9 \% Vol					&	8 \% Vol					\\
\hline
\end{tabular}
\end{center}
\label{tab:results}
\end{table}%

\section{Conclusions}
An equivalent pore fluid in combination with empirical relationships for a porous snow frame model are used to asses the plane wave solutions of Biot's theory for liquid water in snow. 
The results indicate that acoustic waves are sensitive to liquid water content in snow and the phase velocity of the first compressional wave is expected to be lowered by roughly 300~m/s for every increase of 0.1 in water saturation in the pore space. In combination with an independent density measurement acoustic waves should be capable to estimate the liquid water content in snow. 
The use of two end-member models for snow compaction reveal that additional measurements of attenuation of the first compressional wave or the phase velocity of the second compressional wave, should this be possible, would allow to estimate also the density of the snowpack from the acoustic measurements.
A field measurement illustrates the simplicity and robustness of the method and highlights the importance of the precision of the density measurement on the estimate of liquid pore water content.

\section{Acknowledgements}
This research was founded by the Swiss National Science Foundation. The author appreciates the Natural Sciences and Engineering Research Council of Canada (NSERC) for covering the traveling costs and the WSL-Institut f\"ur Schnee- und Lawinenforschung SLF for providing the acoustic measurement device for the field experiment. 

\appendix

\section{Compressional plane wave velocities in a Biot-type porous material}
\label{apx:planewave}

A convenient method to solve Biot's \citeyearpar{biot:1956a} differential equations for wave propagation in porous materials in closed form is to assume plane waves as a solution \citep{biot:1956a,carcione:2007}. Plane waves of the form
\begin{eqnarray}
\vartheta_{S} = \vartheta_{S0} \: e^{i(\omega t -kx)}, \\
\vartheta_{f} = \vartheta_{f0} \: e^{i(\omega t -kx)},
\end{eqnarray}
where $\vartheta_{S}$ and $\vartheta_{f}$ are the velocities of the waves propagating in the solid and fluid, $\omega$ is the angular frequency, $k$ is the wavenumber and $i = \sqrt{-1}$, 
are substituted into Biot's differential equations
\begin{eqnarray}
\label{biot1}
\partial_{j} \sigma_{ij}^{(S)} = \rho_{11} \partial_{tt}^2 u_{i}^{(S)} + \rho_{12} \partial_{tt}^2 u_{i}^{(f)} + b (u_{i}^{(S)} - u_{i}^{(f)}) ,\\
- \phi \partial_j p_f = \rho_{12} \partial_{tt}^2 u_{i}^{(S)} + \rho_{22} \partial_{tt}^2 u_{i}^{(f)} - b (u_{i}^{(S)} - u_{i}^{(f)}) ,
\label{biot2}
\end{eqnarray}
where $\sigma$ is the stress tensor, $\rho_{ij}$ are mass coefficients which take into account that the relative fluid flow through the pores is not uniform, $b = \phi^2 \eta/\kappa$ is the friction coefficient with $\eta$ being the viscosity, $\phi$ the porosity, and $\kappa$ the permeability, $u$ is the displacement vector with the superscripts $(S)$ and $(f)$ referring to the solid matrix and the fluid, respectively. 
Equations \ref{biot1} and \ref{biot2} have to be expressed in the velocity stress formulation to obtain the quadratic dispersion relation
\begin{multline}
\label{eq:dispersion}
- \left( \rho_f^2 + \frac{i}{\omega} Y \rho \right) v_c^4 + \\
 \left[ \frac{i}{\omega} Y \left( K_G + \frac{3}{4} \mu_S \right) + M( 2 \alpha \rho_f - \rho ) \right] v_c^2 + \\
 M \left( K_S + \frac{3}{4} \mu_S \right) = 0 ,
\end{multline}
that can be solved for the two complex plain wave moduli $v_c = \omega / k$ of the first and second compressional wave.
Symbols in equation \ref{eq:dispersion} are the fluid density $\rho_f$, the viscodynamic operator $Y$
\begin{equation}
\label{eq:viscoop}
Y(t) = m \partial_t \delta(t) + \frac{\eta}{\kappa} \delta(t),
\end{equation}
with $\delta(t)$ being the Dirac function and
\begin{equation}
m = \frac{\rho_f \cal{T}}{\phi} ,
\end{equation}
where $\cal{T}$ is the tortuosity. Gassman's \citeyearpar{gassmann:1951} modulus $K_G$ is 
\begin{equation}
K_G = K_m + \alpha^2 M,
\end{equation}
with the effective stress coefficient
\begin{equation}
\alpha = 1- \frac{ K_{\rm m}}{K_{\rm S} } ,
\end{equation}
the poroelastic incompressibility
\begin{equation}
M= {\left[ \frac{(\alpha -\phi )}{K_s}+\frac{\phi }{K_f} \right]}^{-1} ,
\end{equation}
and $K_s$, $K_m$ and $K_f$ being the frame, matrix and fluid bulk moduli.
The phase velocity and attenuation can be obtained from the complex plain wave modulus using
\begin{equation}
\label{phasevel}
V (\omega) = \left[ \mbox{Re}(V_c (\omega) ) ^{-1} \right]^{-1} ,
\end{equation}
and
\begin{equation}
\label{attenuattion}
Q_p (\omega)^{-1} = 2 \frac{\mbox{Im}(V_c (\omega) )}{\mbox{Re}(V_c (\omega) )} ,
\end{equation}
\citep{oconnel:1978}.

\bibliographystyle{model2-names} 
\bibliography{0-references}

\begin{thebibliography}{45}
\expandafter\ifx\csname natexlab\endcsname\relax\def\natexlab#1{#1}\fi
\expandafter\ifx\csname url\endcsname\relax
  \def\url#1{\texttt{#1}}\fi
\expandafter\ifx\csname urlprefix\endcsname\relax\def\urlprefix{URL }\fi
\providecommand{\eprint}[2][]{\url{#2}}
\providecommand{\bibinfo}[2]{#2}
\ifx\xfnm\relax \def\xfnm[#1]{\unskip,\space#1}\fi
\bibitem[{Biot(1956a)}]{biot:1956}
\bibinfo{author}{Biot, M.A.}, \bibinfo{year}{1956}a.
\newblock \bibinfo{title}{Theory of propagation of elastic waves in a
  fluid-saturated porous solid. {I}. {L}ow-frequency range}.
\newblock \bibinfo{journal}{Journal of the Acoustical Society of America}
  \bibinfo{volume}{28}, \bibinfo{pages}{168--178}.
\bibitem[{Biot(1956b)}]{biot:1956a}
\bibinfo{author}{Biot, M.A.}, \bibinfo{year}{1956}b.
\newblock \bibinfo{title}{Theory of propagation of elastic waves in a
  fluid-saturated porous solid. {II. Higher} frequency range}.
\newblock \bibinfo{journal}{Journal of the Acoustical Society of America}
  \bibinfo{volume}{28}, \bibinfo{pages}{179--191}.
\bibitem[{Biot(1962)}]{biot:1962}
\bibinfo{author}{Biot, M.A.}, \bibinfo{year}{1962}.
\newblock \bibinfo{title}{Mechanics of deformation and acoustic propagation in
  porous media}.
\newblock \bibinfo{journal}{Journal of Applied Physics} \bibinfo{volume}{33},
  \bibinfo{pages}{1482--1498}.
\bibitem[{Biot and Willis(1957)}]{biot:1957}
\bibinfo{author}{Biot, M.A.}, \bibinfo{author}{Willis, D.G.},
  \bibinfo{year}{1957}.
\newblock \bibinfo{title}{The elastic coefficients of the theory of
  consolidation}.
\newblock \bibinfo{journal}{Journal of Applied Mechanics} \bibinfo{volume}{24},
  \bibinfo{pages}{594--601}.
\bibitem[{Carcione(2007)}]{carcione:2007}
\bibinfo{author}{Carcione, J.M.}, \bibinfo{year}{2007}.
\newblock \bibinfo{title}{Wave Fields in Real Media: Wave Propagation in
  Anisotropic, Anelastic, Porous and Electromagnetic Media}.
\newblock \bibinfo{publisher}{Elsevier}, \bibinfo{address}{Amsterdam}.
  \bibinfo{edition}{2nd} edition.
\bibitem[{Carcione et~al.(2006)Carcione, Picotti, Gei and
  Rossi}]{carcione:2006b}
\bibinfo{author}{Carcione, J.M.}, \bibinfo{author}{Picotti, S.},
  \bibinfo{author}{Gei, D.}, \bibinfo{author}{Rossi, G.}, \bibinfo{year}{2006}.
\newblock \bibinfo{title}{Physics and seismic modeling for monitoring {CO$_2$}
  storage}.
\newblock \bibinfo{journal}{Pure and Applied Geophysics} \bibinfo{volume}{163},
  \bibinfo{pages}{175--207}.
\bibitem[{Colbeck(1979)}]{colbeck:1979}
\bibinfo{author}{Colbeck, S.C.}, \bibinfo{year}{1979}.
\newblock \bibinfo{title}{Water flow through heterogeneous snow}.
\newblock \bibinfo{journal}{Cold Regions Science and Technology}
  \bibinfo{volume}{1}, \bibinfo{pages}{37--45}.
\bibitem[{Colbeck(1982)}]{colbeck:1982}
\bibinfo{author}{Colbeck, S.C.}, \bibinfo{year}{1982}.
\newblock \bibinfo{title}{An overview of seasonal snow metamorphism}.
\newblock \bibinfo{journal}{Reviews of Geophysics} \bibinfo{volume}{20},
  \bibinfo{pages}{45--61}.
\bibitem[{Conway and Raymond(1993)}]{conway:1993}
\bibinfo{author}{Conway, H.}, \bibinfo{author}{Raymond, C.},
  \bibinfo{year}{1993}.
\newblock \bibinfo{title}{Snow stability during rain}.
\newblock \bibinfo{journal}{Journal oJGlaciology} \bibinfo{volume}{39}.
\bibitem[{Darcy(1856)}]{darcy:1856}
\bibinfo{author}{Darcy, H.}, \bibinfo{year}{1856}.
\newblock \bibinfo{title}{Les fontaines publiques de la ville de {Dijon}}.
\newblock \bibinfo{publisher}{Dalmont}, \bibinfo{address}{Paris}.
\bibitem[{Denoth(1989)}]{denoth:1989}
\bibinfo{author}{Denoth, A.}, \bibinfo{year}{1989}.
\newblock \bibinfo{title}{Snow dielectric measurements}.
\newblock \bibinfo{journal}{Advances in Space Research} \bibinfo{volume}{9},
  \bibinfo{pages}{233--243}.
\bibitem[{Denoth et~al.(1984)Denoth, Foglar, Weiland, M{\"a}tzler, Aebischer,
  Tiuri and Sihvola}]{denoth:1984}
\bibinfo{author}{Denoth, A.}, \bibinfo{author}{Foglar, A.},
  \bibinfo{author}{Weiland, P.}, \bibinfo{author}{M{\"a}tzler, C.},
  \bibinfo{author}{Aebischer, H.}, \bibinfo{author}{Tiuri, M.},
  \bibinfo{author}{Sihvola, A.}, \bibinfo{year}{1984}.
\newblock \bibinfo{title}{A comparative study of instruments for measuring the
  liquid water content of snow}.
\newblock \bibinfo{journal}{Journal of Applied Physics} \bibinfo{volume}{56},
  \bibinfo{pages}{2154--2160}.
\bibitem[{Fierz et~al.(2009)Fierz, Armstrong, Durand, Etchevers, Greene,
  McClung, Nishimura, Satyawali and Sokratov}]{fierz:2009}
\bibinfo{author}{Fierz, C.}, \bibinfo{author}{Armstrong, R.L.},
  \bibinfo{author}{Durand, Y.}, \bibinfo{author}{Etchevers, P.},
  \bibinfo{author}{Greene, E.}, \bibinfo{author}{McClung, D.M.},
  \bibinfo{author}{Nishimura, K.}, \bibinfo{author}{Satyawali, P.K.},
  \bibinfo{author}{Sokratov, S.A.}, \bibinfo{year}{2009}.
\newblock \bibinfo{title}{TheInternational Classificationi for Seasonal Snow on
  the Groun}.
\newblock \bibinfo{publisher}{UNESCO/IHP}.
\bibitem[{Gassmann(1951)}]{gassmann:1951}
\bibinfo{author}{Gassmann, F.}, \bibinfo{year}{1951}.
\newblock \bibinfo{title}{{\"Uber die Elastizit\"at por\"oser Medien}}.
\newblock \bibinfo{journal}{{Vierteljahresschrift der Naturforschenden
  Gesellschaft in Z\"urich}} \bibinfo{volume}{96}, \bibinfo{pages}{1--23}.
\bibitem[{Gerdel(1954)}]{gerdel:1954}
\bibinfo{author}{Gerdel, R.W.}, \bibinfo{year}{1954}.
\newblock \bibinfo{title}{The transmission of water through snow}.
\newblock \bibinfo{journal}{Transactions, American Geophysical Union}
  \bibinfo{volume}{35}, \bibinfo{pages}{475--485}.
\bibitem[{Gold(1958)}]{gold:1958}
\bibinfo{author}{Gold, L.W.}, \bibinfo{year}{1958}.
\newblock \bibinfo{title}{Some observations on the dependence of strain on
  stress for ice}.
\newblock \bibinfo{journal}{Canadian Journal of Physics} \bibinfo{volume}{36},
  \bibinfo{pages}{1265--1275}.
\bibitem[{Gupta et~al.(2005)Gupta, Haritashya and Singh}]{gupta:2005}
\bibinfo{author}{Gupta, R.}, \bibinfo{author}{Haritashya, U.},
  \bibinfo{author}{Singh, P.}, \bibinfo{year}{2005}.
\newblock \bibinfo{title}{Mapping dry/wet snow cover in the {Indian Himalayas}
  using {IRS} multispectral imagery}.
\newblock \bibinfo{journal}{Remote Sensing of Environment}
  \bibinfo{volume}{97}, \bibinfo{pages}{458--469}.
\bibitem[{Ishida(1965)}]{ishida:1965}
\bibinfo{author}{Ishida, T.}, \bibinfo{year}{1965}.
\newblock \bibinfo{title}{Acoustic properties of snow}.
\newblock \bibinfo{journal}{Contributions from the Institute of Low Temperature
  Science} \bibinfo{volume}{20}, \bibinfo{pages}{23--63}.
\bibitem[{Jocker and Smeulders(2009)}]{jocker:2009}
\bibinfo{author}{Jocker, J.}, \bibinfo{author}{Smeulders, D.},
  \bibinfo{year}{2009}.
\newblock \bibinfo{title}{Ultrasonic measurements on poroelastic slabs:
  Determination of reflection and transmission coefficients and processing for
  {Biot} input parameters}.
\newblock \bibinfo{journal}{Ultrasonics} \bibinfo{volume}{49},
  \bibinfo{pages}{319--330}.
\bibitem[{Johnson(1982)}]{johnson:1982}
\bibinfo{author}{Johnson, J.B.}, \bibinfo{year}{1982}.
\newblock \bibinfo{title}{On the application of {Biot}'s theory to acoustic
  wave propagation in snow}.
\newblock \bibinfo{journal}{Cold Regions Science and Technology}
  \bibinfo{volume}{6}, \bibinfo{pages}{49--60}.
\bibitem[{Jordan(1991)}]{jordan:1991}
\bibinfo{author}{Jordan, R.}, \bibinfo{year}{1991}.
\newblock \bibinfo{title}{A one-dimensional temperature model for a snow cover:
  Technical documentation for SNTHERM. 89.}
\newblock \bibinfo{type}{Technical Report}. DTIC Document.
\bibitem[{Lynch-Stieglitz(1994)}]{lynch:1994}
\bibinfo{author}{Lynch-Stieglitz, M.}, \bibinfo{year}{1994}.
\newblock \bibinfo{title}{The development and validation of a simple snow model
  for the giss gcm}.
\newblock \bibinfo{journal}{Journal of Climate} \bibinfo{volume}{7},
  \bibinfo{pages}{1842--1855}.
\bibitem[{Marshall and Koh(2008)}]{marshall:2008}
\bibinfo{author}{Marshall, H.P.}, \bibinfo{author}{Koh, G.},
  \bibinfo{year}{2008}.
\newblock \bibinfo{title}{Fmcw radars for snow research}.
\newblock \bibinfo{journal}{Cold Regions Science and Technology}
  \bibinfo{volume}{52}, \bibinfo{pages}{118--131}.
\bibitem[{Mavko et~al.(1998)Mavko, Mukerji and Dvorkin}]{mavko:1998}
\bibinfo{author}{Mavko, G.}, \bibinfo{author}{Mukerji, T.},
  \bibinfo{author}{Dvorkin, J.}, \bibinfo{year}{1998}.
\newblock \bibinfo{title}{The Rock Physics Handbook: Tools for Seismic Analysis
  in Porous Media}.
\newblock \bibinfo{publisher}{Cambridge University Press}.
\bibitem[{Mavko et~al.(2009)Mavko, Mukerji and Dvorkin}]{mavko:2009}
\bibinfo{author}{Mavko, G.}, \bibinfo{author}{Mukerji, T.},
  \bibinfo{author}{Dvorkin, J.}, \bibinfo{year}{2009}.
\newblock \bibinfo{title}{The Rock Physics Handbook: Tools for Seismic Analysis
  in Porous Media}.
\newblock \bibinfo{publisher}{Cambridge University Press}.
  \bibinfo{edition}{2nd edition} edition.
\bibitem[{Mavko and Nur(1979)}]{mavko:1979}
\bibinfo{author}{Mavko, G.M.}, \bibinfo{author}{Nur, A.}, \bibinfo{year}{1979}.
\newblock \bibinfo{title}{Wave attenuation in partially saturated rocks}.
\newblock \bibinfo{journal}{Geophysics} \bibinfo{volume}{44},
  \bibinfo{pages}{161--178}.
\bibitem[{Mellor(1975)}]{mellor:1975}
\bibinfo{author}{Mellor, M.}, \bibinfo{year}{1975}.
\newblock \bibinfo{title}{A review of basic snow mechanics}, in:
  \bibinfo{booktitle}{The International Symposium on Snow Mechanics,
  Grindelwald, Switzerland}, \bibinfo{organization}{IAHS-AISH}. pp.
  \bibinfo{pages}{251--291}.
\bibitem[{M{\"u}ller et~al.(2010)M{\"u}ller, Gurevich and
  Lebedev}]{muller:2010}
\bibinfo{author}{M{\"u}ller, T.M.}, \bibinfo{author}{Gurevich, B.},
  \bibinfo{author}{Lebedev, M.}, \bibinfo{year}{2010}.
\newblock \bibinfo{title}{Seismic wave attenuation and dispersion resulting
  from wave-induced flow in porous rocks --- {A} review}.
\newblock \bibinfo{journal}{Geophysics} \bibinfo{volume}{75},
  \bibinfo{pages}{75A147--75A164}.
\bibitem[{Nelder and Mead(1965)}]{nelder:1965}
\bibinfo{author}{Nelder, J.A.}, \bibinfo{author}{Mead, R.},
  \bibinfo{year}{1965}.
\newblock \bibinfo{title}{A simplex method for function minimization}.
\newblock \bibinfo{journal}{The computer journal} \bibinfo{volume}{7},
  \bibinfo{pages}{308--313}.
\bibitem[{O'connell and Budiansky(1978)}]{oconnel:1978}
\bibinfo{author}{O'connell, R.}, \bibinfo{author}{Budiansky, B.},
  \bibinfo{year}{1978}.
\newblock \bibinfo{title}{Measures of dissipation in viscoelastic media}.
\newblock \bibinfo{journal}{Geophysical Research Letters} \bibinfo{volume}{5},
  \bibinfo{pages}{5--8}.
\bibitem[{Oura(1952)}]{oura:1952}
\bibinfo{author}{Oura, H.}, \bibinfo{year}{1952}.
\newblock \bibinfo{title}{Reflection of sound at snow surface and mechanism of
  sound propagation in snow}.
\newblock \bibinfo{journal}{Low Temperature Science} \bibinfo{volume}{9},
  \bibinfo{pages}{179--186}.
\bibitem[{Pialucha et~al.(1989)Pialucha, Guyott and Cawley}]{pialucha:1989}
\bibinfo{author}{Pialucha, T.}, \bibinfo{author}{Guyott, C.},
  \bibinfo{author}{Cawley, P.}, \bibinfo{year}{1989}.
\newblock \bibinfo{title}{￼amplitude spectrum method for the measurement of
  phase velocity}.
\newblock \bibinfo{journal}{Ultrasonics} \bibinfo{volume}{27},
  \bibinfo{pages}{270--279}.
\bibitem[{Pride(2005)}]{pride:2005}
\bibinfo{author}{Pride, S.R.}, \bibinfo{year}{2005}.
\newblock \bibinfo{title}{Hydrogeophysics}. \bibinfo{publisher}{Springer}.
  chapter \bibinfo{chapter}{Relationships between seismic and hydrological
  properties}.
\newblock pp. \bibinfo{pages}{253--291}.
\bibitem[{Pride and Berryman(2003a)}]{pride:2003}
\bibinfo{author}{Pride, S.R.}, \bibinfo{author}{Berryman, J.G.},
  \bibinfo{year}{2003}a.
\newblock \bibinfo{title}{Linear dynamics of double-porosity dual-permeability
  materials. i. governing equations and acoustic attenuation}.
\newblock \bibinfo{journal}{Physical Review E} \bibinfo{volume}{68},
  \bibinfo{pages}{036603}.
\bibitem[{Pride and Berryman(2003b)}]{pride:2003a}
\bibinfo{author}{Pride, S.R.}, \bibinfo{author}{Berryman, J.G.},
  \bibinfo{year}{2003}b.
\newblock \bibinfo{title}{Linear dynamics of double-porosity dual-permeability
  materials. ii. fluid transport equations}.
\newblock \bibinfo{journal}{Physical Review E} \bibinfo{volume}{68},
  \bibinfo{pages}{036604}.
\bibitem[{Reine et~al.(2009)Reine, van~der Baan and Clark}]{reine:2009}
\bibinfo{author}{Reine, C.}, \bibinfo{author}{van~der Baan, M.},
  \bibinfo{author}{Clark, R.}, \bibinfo{year}{2009}.
\newblock \bibinfo{title}{The robustness of seismic attenuation measurements
  using fixed- and variable-window time-frequency transforms}.
\newblock \bibinfo{journal}{Geophysics} \bibinfo{volume}{74},
  \bibinfo{pages}{WA123--WA135}.
\bibitem[{Sachse and Pao(1978)}]{sachse:1978}
\bibinfo{author}{Sachse, W.}, \bibinfo{author}{Pao, Y.H.},
  \bibinfo{year}{1978}.
\newblock \bibinfo{title}{On the determination of phase and group velocities of
  dispersive waves in solids}.
\newblock \bibinfo{journal}{Journal of Applied Physics} \bibinfo{volume}{49},
  \bibinfo{pages}{4320--4327}.
\bibitem[{Schweizer et~al.(2003)Schweizer, Jamieson and
  Schneebeli}]{schweizer:2003}
\bibinfo{author}{Schweizer, J.}, \bibinfo{author}{Jamieson, J.B.},
  \bibinfo{author}{Schneebeli, M.}, \bibinfo{year}{2003}.
\newblock \bibinfo{title}{Snow avalanche formation}.
\newblock \bibinfo{journal}{Reviews of Geophysics} \bibinfo{volume}{41},
  \bibinfo{pages}{1--25}.
\bibitem[{Shi and Dozier(1995)}]{shi:1995}
\bibinfo{author}{Shi, J.}, \bibinfo{author}{Dozier, J.}, \bibinfo{year}{1995}.
\newblock \bibinfo{title}{Inferring snow wetness using c-band data from sir-c's
  polarimetric synthetic aperture radar}.
\newblock \bibinfo{journal}{Geoscience and Remote Sensing, IEEE Transactions
  on} \bibinfo{volume}{33}, \bibinfo{pages}{905--914}.
\bibitem[{Sidler(2014)}]{sidler:2014a}
\bibinfo{author}{Sidler, R.}, \bibinfo{year}{2014}.
\newblock \bibinfo{title}{Snow slab failure due to {Biot-type} acoustic wave
  propagation}, in: \bibinfo{booktitle}{Proceedings ISSW}, pp.
  \bibinfo{pages}{146 -- 150}.
\bibitem[{Sommerfeld(1982)}]{sommerfeld:1982}
\bibinfo{author}{Sommerfeld, R.}, \bibinfo{year}{1982}.
\newblock \bibinfo{title}{A review of snow acoustics}.
\newblock \bibinfo{journal}{Reviews of Geophysics} \bibinfo{volume}{20},
  \bibinfo{pages}{62--66}.
\bibitem[{Tang and Cheng(2004)}]{tang:2004}
\bibinfo{author}{Tang, X.M.}, \bibinfo{author}{Cheng, A.},
  \bibinfo{year}{2004}.
\newblock \bibinfo{title}{Quantitative Borehole Acoustic Methods}.
\newblock Handbook of Geophysical Exploration, \bibinfo{publisher}{Elsevier},
  \bibinfo{address}{Amsterdam}.
\bibitem[{Techel and Pielmeier(2011)}]{techel:2011}
\bibinfo{author}{Techel, F.}, \bibinfo{author}{Pielmeier, C.},
  \bibinfo{year}{2011}.
\newblock \bibinfo{title}{Point observations of liquid water content in wet
  snow--investigating methodical, spatial and temporal aspects}.
\newblock \bibinfo{journal}{The Cryosphere} \bibinfo{volume}{5},
  \bibinfo{pages}{405--418}.
\bibitem[{Teja and Rice(1981)}]{teja:1981}
\bibinfo{author}{Teja, A.}, \bibinfo{author}{Rice, P.}, \bibinfo{year}{1981}.
\newblock \bibinfo{title}{Generalized corresponding states method for the
  viscosities of liquid mixtures}.
\newblock \bibinfo{journal}{Industrial \& Engineering Chemistry Fundamentals}
  \bibinfo{volume}{20}, \bibinfo{pages}{77--81}.
\bibitem[{Wood(1955)}]{wood:1955}
\bibinfo{author}{Wood, A.B.}, \bibinfo{year}{1955}.
\newblock \bibinfo{title}{A textbook of sound: The physics of vibration}.
\newblock \bibinfo{publisher}{Bell and Sons}.

\end{thebibliography}

\end{document}